\documentclass[letterpaper,12pt]{article}
\usepackage[utf8]{inputenc}
\DeclareUnicodeCharacter{211D}{\ensuremath{\mathbb{R}}}
\usepackage{setspace}
\usepackage[margin=1in]{geometry}
\usepackage{authblk}
\usepackage[round]{natbib}
\usepackage[shortlabels]{enumitem}
\usepackage{amssymb}
\usepackage{amsmath}
\usepackage{amsthm}
\usepackage{mathtools}
\usepackage{bbm}
\usepackage{textcomp, gensymb}
\usepackage{url}
\usepackage[colorlinks,linkcolor=blue,citecolor=blue,urlcolor=blue,filecolor=blue]{hyperref}
\usepackage[parfill]{parskip}
\usepackage{graphicx}
\usepackage{comment}
\usepackage{float}
\usepackage{multirow}
\usepackage[font=small]{caption}
\usepackage{appendix}
\floatstyle{plaintop}
\restylefloat{table}
\newcommand{\bld}[1]{\boldsymbol{\mathbf{#1}}}
\DeclareMathOperator{\Tr}{Tr}
\numberwithin{equation}{section}
\numberwithin{figure}{section}
\numberwithin{table}{section}
\newtheorem{proposition}{Proposition}
\newtheorem{corollary}{Corollary}[proposition]

\title{Markov Random Fields with Proximity Constraints for Spatial Data}
\author[1]{Sudipto Saha}
\author[1]{Jonathan R. Bradley}
\affil[1]{Department of Statistics, Florida State University, Tallahassee, FL 32306, USA.}

\date{}

\begin{document}
\maketitle{}

\begin{abstract}
    The conditional autoregressive (CAR) model, simultaneous autoregressive (SAR) model, and its variants have become the predominant strategies for modeling regional or areal-referenced spatial data. The overwhelming wide-use of the CAR/SAR model motivates the need for new classes of models for areal-referenced data. Thus, we develop a novel class of Markov random fields based on truncating the full-conditional distribution. We define this truncation in two ways leading to versions of what we call the truncated autoregressive (TAR) model. First, we truncate the full conditional distribution so that a response at one location is close to the average of its neighbors. This strategy establishes relationships between TAR and CAR. Second, we truncate on the joint distribution of the data process in a similar way. This specification leads to connection between TAR and SAR model. Our Bayesian implementation does not use Markov chain Monte Carlo (MCMC) for Bayesian computation, and generates samples directly from the posterior distribution. Moreover, TAR does not have a range parameter that arises in the CAR/SAR models, which can be difficult to learn. We present the results of the proposed truncated autoregressive model on several simulated datasets and on a dataset of average property prices.
    
\end{abstract}

\section{Introduction} \label{2sec:intro}
The key idea of a Markov random field (MRF; introduced by \citealp{dobruschin1968description}) is that the conditional density of one location is dependent only on its neighborhood locations \citep{spitzer1971markov,besag1974spatial}, and this type of specification is considered to be more ``local'' \citep{banerjee2015hierarchical}. \citet{besag1974spatial} developed a popular Markov random field, the conditional autoregressive (CAR) model, which defines the joint distribution to be a Gibbs distribution \citep{geman1984stochastic} through the use of \textit{Brook's Lemma} \citep{brook1964distinction}, and an explicable set of full-conditional distributions. CAR model possesses a computational advantage over directly specifying covariance function. Since the precision matrix is defined directly in case of a CAR model, no inversion of covariance matrix is needed while performing inferences. The CAR model introduces an aforementioned range parameter to ensure a positive definite covariance matrix. Markov random fields have been consistently and widely used in spatial statistics (see \citealp{besag1974spatial,cressie1993statistics,cressie2011statistics,sorbye2014scaling,banerjee2015hierarchical}, among several others).

The importance of the CAR model \citep{besag1974spatial,besag1986statistical,besag1991bayesian} for the analysis of areal-referenced data cannot be overstated. At the time of writing this article a non-exhaustive list of recent papers that apply the CAR or some closely related variant include: \citet{bradley2024generating,galeana2024impact,he2024incorporating,kang2024fast,leveau2024spatial,sakamoto2024impacts,sasmita2024spatial,st2024high,wang2024using,zhuang2024neighborhood}. Moreover, the key early references for the CAR including \citet{besag1974spatial}, \citet{besag1986statistical}, and \citet{besag1991bayesian}, currently have over 20,000 citations demonstrating its importance to the scientific literature. Outside of the CAR model there are fewer alternatives to modeling areal data as compared to the myriad of options in the point-referenced setting (e.g., see \citealp{heaton2019case} for a comparison of a long incomplete list of methods in the point referenced setting). The CAR’s wide use should not give reason to overlook common issues that arise with CAR models. For example, to ensure the CAR model is proper one needs to introduce a range parameter to make the implied spatial covariance matrix positive definite \citep{cressie1993statistics,banerjee2015hierarchical}, and the parameter space for this range parameter can be difficult to interpret. CAR models also have limited flexibility in modeling dependence, requiring known proximity matrices. Several have considered unknown adjacency structure that have limited correlation structure \citep[e.g., see][]{white2009stochastic,ma2010hierarchical}, but these models can be exceedingly difficult to implement in practice \citep{gao2019bayesian}. Additionally, many have argued that the known proximity matrix with a single range parameter in CAR models insufficiently describes spatial correlations \citep{rodrigues2012bayesian}. In a similar vain, others have found that the intrinsic conditional autoregressive model, which forces the range parameter to be 1, only allows for short-range correlations \citep{lawson2015shared} that may not be realistic. Thus, the goal of this article is to explore a new class of MRFs that avoid the need for a range parameter.

We propose a class of MRFs that truncate the statistical model based on proximity constraints that again results in an expression of the precision matrix, but avoids the introduction of a range parameter to ensure a positive definite precision matrix. This modeling strategy requires methodologists to define two quantities, namely a local average, and a prior for the bound that constrains a response-value to be close to the local average. This gives an avenue for methodologists to consider several types of prior distributions and constraining around their favorite local average/predictor for future in development. In this article, we start with a simple equally weighted local average and a type of uniform prior on the bound, and consider truncating both the full-conditional distribution and the joint distribution of the responses. These specifications lead to what we call the conditional truncated autoregressive (TAR\textsubscript{C}) model and simultaneous truncated autoregressive (TAR\textsubscript{S}) model respectively. The TAR\textsubscript{C} and CAR model, and the TAR\textsubscript{S} and SAR model have interesting relationships that provide insights to how this general strategy to truncate can be interpreted according to these classical models. What is particularly useful about the TAR model is that the parameters include a regression effect, and two variances. Hence, it avoids the known difficulties that arise with the spatial range parameter.

The usefulness of truncated Gaussian random field to analyze truncated spatial data was first introduced by \citet{stein1992prediction}, where he was solving the problem of modeling sparse spatial data. \citet{sanso2004bayesian} also used a truncated normal model, like \citet{stein1992prediction}, to analyze rainfall data which is naturally truncated for negative values. This strategy to enforce dependence by artificially truncating a distribution based on proximity constraints has not been considered directly (e.g., see \citealp{bradley2021empirical}, and \citealp{zong2023criterion}, for different artificially constrained models), and is clearly different from the standard strategies of directly specifying the covariance function/spectral density \citep{stein1999interpolation} and defining shared random effects as coefficients in a basis function expansion (see \citealp{watson1984smoothing,wahba1990spline,nychka2002multiresolution,banerjee2008gaussian,cressie2008fixed}, among several others). Consequently, this particular direction to enforce dependence is an area that has been consistently overlooked, and partially motivates our novel development.

Bayesian implementation of our consideration of TAR can be achieved without MCMC, leading to efficient sampling that avoids tuning and MCMC diagnostics. There has been renewed interest in developing strategies for sampling directly from the posterior distribution \citep{zhang2019practical,banerjee2020conjugate,bradley2023deep,bradley2024generating}. Moreover, the precision structure also lends itself naturally to the use of the Neumann series \citep{ravishanker2021first} leading to fast prediction via posterior predictive step. 

The remainder of the article is organized as follows. In Section \ref{2sec:review}, we provide a brief review of the CAR and SAR model. In Section \ref{2sec:motivation}, we provide a review of constrained spatial models along with discussing the motivation behind proposing the truncated autoregressive model. In Section \ref{2sec:method}, we present the proposed TAR model. Specifically, we define the proposed model in various ways that helps to connect the model to CAR and SAR, and we present several results to establish the theoretical foundation of the TAR model. We also show the implementation of the model using a sampler to directly sample from the posterior distribution. In Section \ref{2sec:simulation}, we demonstrate the performance of our model by providing several illustrations through multiple simulated datasets, followed by a more extensive simulation study to compare the performance of our model to the traditional models. We demonstrate that the TAR model outperforms the traditional models when the data is generated form the TAR model itself, and it shows competitive performance even when the data is generated from the traditional models. We also show that our model is fast. In Section \ref{2sec:application}, we apply our model to a property prices dataset with average prices across Greater Glasgow, Scotland in 2008. Finally, a concluding discussion is presented in Section \ref{2sec:discussion}. We provide the proofs of all the technical results in the Appendix for convenience.

\section{Review} \label{2sec:review}
We review two traditional areal models -- (a) conditional autoregressive (CAR) model, and (b) simultaneous autoregressive (SAR) model.

\subsection{A Review of Conditional Autoregressive Model} \label{2subsec:reviewCAR}
MRFs for areal data are defined by assuming conditional independence between the $i$-th response and all non-neighboring response values given the $i$-th response value's neighbors. Conditional autoregressive (CAR; see \citealp{besag1974spatial,cressie1993statistics,banerjee2015hierarchical} for standard references) is a commonly used Markov random field model that adopts a particular Gaussian assumption. Let us suppose, $\bld{y} \equiv (y_{1},\ldots,y_{n})'$ where $y_{i}$ is the data observed at region $i$. Then the full-conditional distribution of CAR model is given by
\begin{equation} \label{2eq:carCond}
    f(y_{i}\vert y_{-i})=f_{CAR}(y_{i}\vert \bld{y}(N_{i})) = \mathcal{N}\left(\sum_{j\in N_{i}} a_{ij}y_{j},\tau_{i}^{2}\right), \forall i=1,\ldots,n,
\end{equation}
where $y_{-i}=(y_{1},\ldots,y_{i-1},y_{i+1},\ldots,y_{n})$, $N_{i}=\{j:i \neq j,j\sim i\}$, and $f(\cdot)$ is used for distributions of areal data. We use the notation $j \sim i$ to indicate that regions $i$ and $j$ are neighbors. Through \textit{Brook's Lemma} \citep{brook1964distinction,besag1974spatial} one can obtain
\begin{equation} \label{2eq:carJoint1}
    f(y_{1},\ldots,y_{n}) \propto \exp\left(-\frac{1}{2}\bld{y}'\bld{D}^{-1}(\bld{I}-\bld{A})\bld{y}\right),
\end{equation}
where the $n \times n$ matrix $\bld{A}=(a_{ij})$ and $\bld{D}$ is an $n \times n$ diagonal matrix with $D_{ii}=\tau_i^{2}$. Often, $a_{ij}$ is set to $a_{ij}=w_{ij}/\sum_j w_{ij}$ and $\tau_i^{2}$ is set to $\tau_i^{2}=\tau^{2}/\sum_j w_{ij}$, where $w_{ij}$ takes a value of 1 when $j$ is a neighbor of $i$, otherwise it takes a value of 0. The $n \times n$ matrix $\bld{W}=(w_{ij})$ is known as the proximity matrix. In this case, Equation \eqref{2eq:carJoint1} is written as
\begin{equation} \label{2eq:carJoint2}
    f(y_{1},\ldots,y_{n}) \propto \exp\left(-\frac{1}{2\tau^{2}}\bld{y}'(\bld{D}_w-\bld{W})\bld{y}\right),
\end{equation}
where $\bld{D}_w$ is a diagonal matrix with $(D_w)_{ii}=\sum_j w_{ij}$. However, the precision matrix $\frac{1}{\tau^{2}}(\bld{D}_w-\bld{W})$ is not strictly positive definite which makes the joint distribution in \eqref{2eq:carJoint2} an improper distribution. To overcome this issue, $\bld{W}$ is often replaced with $\rho\bld{W}$ and $\rho$ is chosen in a way that $\frac{1}{\tau^{2}}(\bld{D}_w-\rho\bld{W})$ becomes strictly positive definite. Particularly, $\rho$ is restricted to $\left(1/\lambda_{(1)},1/\lambda_{(n)}\right)$, where $\lambda_{(1)},\ldots,\lambda_{(n)}$ are ordered eigenvalues of $\bld{D}_{w}^{-1/2}\bld{W}\bld{D}_{w}^{-1/2}$. A detailed discussion can be found in \citet{banerjee2015hierarchical}.

\subsection{A Review of Simultaneous Autoregressive Model} \label{2subsec:reviewSAR}
The system of random variables in Equation \eqref{2eq:carCond} can be written as $\bld{y}=\bld{Ay}+\bld{\epsilon}$, where $\bld{\epsilon} \sim \mathcal{N}(\bld{0},\bld{D})$ with $\bld{D}$ now being a diagonal matrix where $D_{ii}=\sigma_{i}^{2}$. This implies $(\bld{I}-\bld{A})\bld{y}=\bld{\epsilon}$ and hence the distribution of $\bld{y}$ becomes
\begin{equation} \label{2eq:sarJoint1}
    \bld{y} \sim \mathcal{N}\left(\bld{0},(\bld{I}-\bld{A})^{-1}\bld{D}\left((\bld{I}-\bld{A})^{-1}\right)'\right).
\end{equation}
Equation \eqref{2eq:sarJoint1} is a simultaneous autoregressive (SAR) model \citep{cressie1993statistics,banerjee2015hierarchical} where all the elements of $\bld{y}$ and $\bld{A}$ are modeled simultaneously/jointly through $\bld{y}=\bld{Ay}+\bld{\epsilon}$ as opposed to working with the full-conditional distributions. Now, if we set $\bld{D}=\sigma^{2}\bld{I}$, Equation \eqref{2eq:sarJoint1} simplifies to
\begin{equation} \label{2eq:sarJoint2}
    \bld{y} \sim \mathcal{N}\left(\bld{0},\sigma^{2}\left[(\bld{I}-\bld{A})'(\bld{I}-\bld{A})\right]^{-1}\right).
\end{equation}
Notice that, $(\bld{I}-\bld{A})$ must be full rank for the distribution in \eqref{2eq:sarJoint2} to be proper. To make this happen, a common choice is replacing $\bld{A}$ with $\rho\bld{A}$. In this case, $\rho$ is restricted to have specific values to make the distribution of $\bld{y}$ proper. A detailed discussion regarding this can be found in \citet{banerjee2015hierarchical}, pg. 85. The SAR model can be written vaguely as a CAR, but not every CAR is a SAR model. As such the SAR model can be seen as a special case of the CAR \citep[see][]{verhoef2018relationship}.

\section{Motivation} \label{2sec:motivation}
A less common strategy to include spatial dependence into a model is a constrained spatial model. Typically this model is used when truncation naturally arises in the data. To analyze such type of dataset, truncated Gaussian random field is often used. The probability density function (pdf) of the truncated Gaussian distribution is given by
\begin{equation} \label{2eq:truncatedNormal}
    f(\bld{x};\bld{\mu},\bld{\Sigma},S)=\frac{1}{p(S)}\exp \left(-\frac{1}{2}(\bld{x}-\bld{\mu})'\bld{\Sigma}^{-1}(\bld{x}-\bld{\mu})\right) \mathbbm{1} (\bld{x} \in S),
\end{equation}
where $\bld{x} \in \mathbb{R}^{n}$, $\bld{\mu} \in \mathbb{R}^{n}$, $S \in \mathbb{R}^{n}$, $\bld{\Sigma}$ is an $n \times n$ positive definite matrix, $\mathbbm{1}(\cdot)$ is the indicator function, and $p(S)=\int_{S} \exp \left(-\frac{1}{2}(\bld{x}-\bld{\mu})'\bld{\Sigma}^{-1}(\bld{x}-\bld{\mu})\right)d\bld{x}>0$. The usefulness of truncated Gaussian random field to analyze truncated spatial data was first introduced by \citet{stein1992prediction} for modeling sparse spatial data.

Our motivation to propose a TAR model model comes from the realization that even when the data is not naturally truncated and $\bld{\Sigma}$ in \eqref{2eq:truncatedNormal} is a diagonal matrix (i.e., covariance is zero), the implied covariance is non-zero for neighborhood locations upon imposing the spatial dependence through truncation $S$ in \eqref{2eq:truncatedNormal}. To verify this, we simulate 1000 values of $\bld{x}$ from \eqref{2eq:truncatedNormal} with $n=100$, where each time we set $\bld{\mu}=\bld{0}_{100}$ ($\bld{0}_{100}$ is a 100-dimensional vector of zeros), $\bld{\Sigma}=\bld{I}_{100}$ ($\bld{I}_{100}$ is a $100 \times 100$ identity matrix), and we include proximity information in $S$ as
\begin{equation} \label{2eq:S}
    S = \left\{(\bld{x}_{1},\ldots,\bld{x}_{1000}):(\bld{e}_i'\bld{x}-\bld{e}_i'\bld{Bx})'(\bld{e}_i'\bld{x}-\bld{e}_i'\bld{Bx})<k,\forall i=1,\ldots,100\right\},
\end{equation}
where $\bld{B} \in [0,1]^{100 \times 100}$ is a weighted proximity matrix, $k$ is a strictly positive real number that ensures $\bld{x}$ to be close to its local average $\bld{Bx}$, and $\bld{e}_i$ is a 100-dimensional vector of zeros with only $i$-th element as 1.

\begin{figure}[t!]
    \centering
    \includegraphics[scale=0.8]{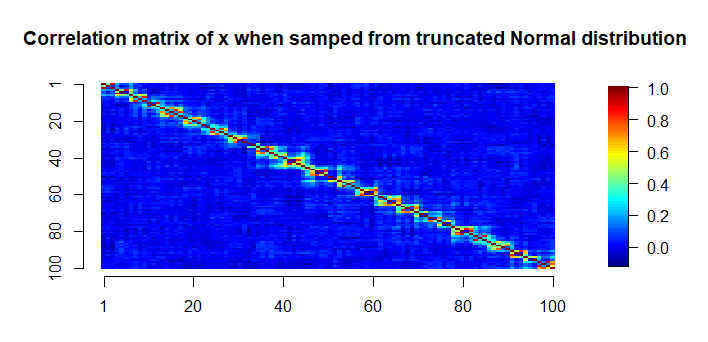}
    \caption{The correlation matrix of $\bld{x} \in \mathbb{R}^{100}$ based on 1000 replicates from \eqref{2eq:truncatedNormal}, where $\bld{\mu}=\bld{0}_{100}$, $\bld{\Sigma}=\bld{I}_{100}$, and $S$ is based on \eqref{2eq:S}.}
    \label{2fig:corplot}
\end{figure}

We calculate the empirical correlation matrix of $\bld{x}$ from those 1000 values which is shown in Figure \ref{2fig:corplot}. Clearly, non-zero correlations show up even when $\bld{\Sigma}=\bld{I}$ in \eqref{2eq:truncatedNormal}, and these non-zero correlations appear because of the proximity information enforced through the truncation. This allows one to parameterize a statistical model based on distance/proximity weighting (e.g., $\bld{Bx}$) \citep{cressie1993statistics}. Since the locations are sorted in increasing order, the neighborhood locations appear around the diagonal. By the definition of $\bld{S}$ in \eqref{2eq:S}, $\bld{x}$ is enforced to be close to its local average which results in non-zero correlations at the neighborhood locations, and this verifies our argument. The novelty of this strategy is enforcing spatial dependence by defining a valid Markov random field through spatial constraints.

\section{Methodology} \label{2sec:method}
In this section, we propose the TAR model and describe several connections to traditional CAR and SAR models. We describe the construction of TAR via conditional distribution in Section \ref{2subsec:ConditionalTrunc}, and through the joint distribution distribution of $\bld{y}$ in Section \ref{2subsec:JointTrunc}. We also explore the TAR\textsubscript{C}/CAR and TAR\textsubscript{S}/SAR relationship. We end this section with Bayesian model implementation (Section \ref{2subsec:BHM}) and spatial prediction (Section \ref{2subsec:postpred}).

\subsection{Truncation on Conditional Distribution} \label{2subsec:ConditionalTrunc}
Let, $\bld{X}\in \mathbb{R}^{n \times p}$ with the $p$-dimensional $i$-th row of $\bld{X}$ denoted with $\bld{x}_{i}'$. Consider the unnormalized truncated full-conditional distribution $\forall i=1,\ldots,n$,
\begin{align}
    f\Bigl(Y_{i}\vert \bld{y}(N_{i}),\bld{\beta},\sigma_{Yi}^{2},\tau_{Yi}^{2},u_{i}\Bigr) \propto &\frac{1}{\sqrt{2\pi \tau_{Yi}^{2}}} \exp\left(-\frac{\widetilde{Y}_{i}^{2}}{2\tau_{Yi}^{2}}\right) \nonumber \\
    &\mathbbm{1}\left(- \sqrt{-2\sigma_{Yi}^{2}\log u_{i}} < \widetilde{Y}_{i} - \sum_{j \in N_{i}} a_{ij}\widetilde{Y}_{j} < \sqrt{-2\sigma_{Yi}^{2}\log u_{i}}\right), \label{2eq:trunc1}
\end{align}
where the mean adjusted $Y_{i}$ is $\widetilde{Y}_{i}=Y_{i}-\bld{x}_{i}'\bld{\beta}$, $u_{i}$ is given a uniform prior on $(0,1)$ denoted with $\mathcal{U}(0,1)$, $\forall i=1,\ldots,n$, $N_{i}=\{j:i \neq j,j\sim i\}$, $\sigma_{Yi}^{2}$ and $\tau_{Yi}^{2}$ are variance parameters, and $a_{ij} = w_{ij}/\sum_{j} w_{ij} \in [0,1]$ are weights. Here, $w_{ij}$ takes a value of 1 only when $j \in N_{i}$, else it takes a value of 0. The $n \times n$ dimensional matrix $\bld{W}=(w_{ij})$ is known as \textit{proximity matrix} with elements ones and zeros. Since, we normalize $w_{ij}$ by $\sum_{j} w_{ij}$, we call the $n \times n$ dimensional matrix $\bld{A}=(a_{ij})$ a \textit{row-normalized proximity matrix}. We call the model in \eqref{2eq:trunc1} a ``\textit{conditional truncated autoregressive model}'' (TAR\textsubscript{C}). Notice that, for each $i$, $w_{ij}$, and hence $a_{ij}$, are non-zero only when $j \in N_{i}$. Therefore, the term $\sum_{j \in N_{i}} a_{ij}\widetilde{Y}_{j}$ represents weighted average of the neighborhoods of $i$. So, through the truncation in \eqref{2eq:trunc1} we enforce $\widetilde{Y}_{i}$ to be close to the local average of its neighborhoods by assigning a uniform prior to the auxiliary variable $u_{i}$. One might interpret the constraint $\{u_{1},\ldots,u_{n}\}$ as an auxiliary variable, which is a common strategy in Bayesian models \citep{albert1993bayesian,besag1993spatial,damlen1999gibbs}. Ideally, we would want the values of $u_{i},\forall i=1,\ldots,n$ to be close to 1 so that the bounds on $\widetilde{Y}_{i}$ are small. This interpretation of $\{u_{i}\}$ leads to the following result.

\begin{proposition} \label{2prop:1}
    Upon integrating out $\{u_{i}\}$, the model in \eqref{2eq:trunc1} becomes
    \begin{equation} \label{2eq:condDist}
        f\Bigl(Y_{i}\vert \bld{y}(N_{i}),\bld{\beta},\sigma_{Yi}^{2},\tau_{Yi}^{2}\Bigr) \propto \exp\left(-\frac{\widetilde{Y}_{i}^{2}}{2\tau_{Yi}^{2}} - \frac{\left[\widetilde{Y}_{i} - \sum_{j \in N_{i}} a_{ij}\widetilde{Y}_{j}\right]^{2}}{2\sigma_{Yi}^{2}}\right),
    \end{equation}
    where $\widetilde{Y}_{i}=Y_{i}-\bld{x}_{i}'\bld{\beta}$ for $i=1,\ldots,n$.
\end{proposition}
\textit{Proof:} See Appendix \ref{2app:a1}.

Proposition \ref{2prop:1} is important, where upon comparing to Equation \eqref{2eq:carJoint2} we can start to see the relationship with the CAR model. That is, when $\tau_{Yi}^{2}=\infty$ we obtain the full-conditional distribution of the CAR model adjusted for covariates. We now derive the joint distribution of $Y_{i},\forall i=1,\ldots,n$ given $\bld{\beta}$ and the variance parameters. To do so, we apply Brook's Lemma \citep{brook1964distinction,besag1974spatial} on Equation \eqref{2eq:condDist} and we get a distribution with the precision matrix $\bld{D}_{\tau}^{-1} + \bld{D}_{\sigma}^{-1}(\bld{I}-\bld{A})$, where $\bld{D}_{\tau}$ and $\bld{D}_{\sigma}$ are diagonal matrices with $(D_{\tau})_{ii}=\tau_{Yi}^{2}$ and $(D_{\sigma})_{ii}=\sigma_{Yi}^{2}$ respectively. Now, since the precision matrix is not symmetric (as $\bld{A}$ is not necessarily symmetric), we set $\tau_{Yi}^{2}=\tau_{Y}^{2}/\sum_{j} w_{ij}$ and $\sigma_{Yi}^{2}=\sigma_{Y}^{2}/\sum_{j} w_{ij}$ to make it symmetric (this is a common practice and is mentioned in Section \ref{2subsec:reviewCAR}; see \citealp{banerjee2015hierarchical} for standard references). This process leads to the following proposition.

\begin{proposition} \label{2prop:2}
    From the model in \eqref{2eq:condDist}, the joint distribution of $Y_{i},\forall i=1,\ldots,n$ given $\bld{\beta}$, $\sigma_{Y}^{2}$ and $\tau_{Y}^{2}$ becomes
    \begin{equation} \label{2eq:condY1}
        \bld{y}\vert \bld{\beta},\sigma_{Y}^{2},\tau_{Y}^{2} \sim \mathcal{N}\left(\bld{X\beta},\left[\frac{1}{\tau_{Y}^{2}}\bld{D}_w + \frac{1}{\sigma_{Y}^{2}}(\bld{D}_w-\bld{W})\right]^{-1}\right),
    \end{equation}
    where $\bld{W}$ is proximity matrix and $\bld{D}_w$ is a diagonal matrix with $\left(D_{w}\right)_{ii}=\sum_{j} w_{ij}$.
\end{proposition}
\textit{Proof:} See Appendix \ref{2app:a1}.

Notice that, unlike in the classical model where the nugget effect is added in the covariance matrix \citep[see][for standard references]{cressie1993statistics,banerjee2015hierarchical}, here in \eqref{2eq:condY1} $\frac{1}{\tau_{Y}^{2}}\bld{D}_w$ is added in the precision matrix. We call the parameter $\tau_{Y}^{2}$ a ``\textit{precision-nugget}'' in the sense that we are allowing for a discontinuity in the precision matrix at the diagonal. The precision-nugget leads to two corollaries from Proposition \ref{2prop:2} that are immediate.
\begin{corollary} \label{2cor:1}
    When $\tau_{Y}^{2} = \infty$, the model in \eqref{2eq:condY1} becomes a traditional CAR model with covariance matrix $\sigma_{Y}^{2}(\bld{D}_w-\bld{W})^{-1}$.
\end{corollary}
\textit{Proof:} If we set $\tau_{Y}^{2} = \infty$ into \eqref{2eq:condY1}, it becomes $\bld{y}\vert \bld{\beta},\sigma_{Y}^{2} \sim \mathcal{N}\left(\bld{X\beta},\sigma_{Y}^{2}(\bld{D}_w-\bld{W})^{-1}\right)$. This is a CAR model with mean $\bld{X\beta}$ and covariance matrix $\sigma_{Y}^{2}(\bld{D}_w-\bld{W})^{-1}$ (refer to Section \ref{2subsec:reviewCAR}; also see \citealp{cressie1993statistics,banerjee2015hierarchical} for standard references).

\begin{corollary} \label{2cor:2}
    The precision matrix $\frac{1}{\tau_{Y}^{2}}\bld{D}_w + \frac{1}{\sigma_{Y}^{2}}(\bld{D}_w-\bld{W})$ of the distribution in \eqref{2eq:condY1} is strictly positive definite.
\end{corollary}
\textit{Proof:} We calculate $\bld{v}'\left[\frac{1}{\tau_{Y}^{2}}\bld{D}_w + \frac{1}{\sigma_{Y}^{2}}(\bld{D}_w-\bld{W})\right]\bld{v}=\frac{1}{\tau_{Y}^{2}}\bld{v}'\bld{D}_w\bld{v} + \frac{1}{\sigma_{Y}^{2}}\bld{v}'(\bld{D}_w-\bld{W})\bld{v}$ for any non-zero vector $\bld{v}$. Notice that, $(\bld{D}_w-\bld{W})\bld{v}=\bld{0}$ for $\bld{v}=\bld{1}$. Also, $\bld{v}'(\bld{D}_w-\bld{W})\bld{v}=\bld{v}'\bld{D}_w(\bld{I}-\bld{A})\bld{v}$, where we write $\bld{A}=\bld{D}_{w}^{-1}\bld{W}$ since $a_{ij}=w_{ij}/\sum_{j} w_{ij}$. Notice that for $\bld{Z}=\bld{I}-\bld{A}$, $z_{ii}=1$ for all $i=1,\ldots,n$ and $\sum_{j \neq i} \lvert z_{ij} \rvert=1$ for each $i=1,\ldots,n$. Applying Gershgorin circle theorem \citep{gershgorin1931uber,varga2004geršgorin} on $\bld{Z}$, every eigenvalue of $\bld{Z}$ lies within $[0,2]$. Hence, $\bld{v}'\bld{D}_{w}(\bld{I}-\bld{A})\bld{v}$ is strictly non-negative for any non-zero vector $\bld{v}$. Therefore, $\bld{v}'(\bld{D}_w-\bld{W})\bld{v} \geq 0$. Also, $\bld{v}'\bld{D}_{w}\bld{v} = \sum_{i}v_{i}^{2}\sum_{j} w_{ij} > 0$. Therefore, $\bld{v}'\left[\frac{1}{\tau_{Y}^{2}}\bld{D}_w + \frac{1}{\sigma_{Y}^{2}}(\bld{D}_w-\bld{W})\right]\bld{v} > 0$. Hence, $\frac{1}{\tau_{Y}^{2}}\bld{D}_w + \frac{1}{\sigma_{Y}^{2}}(\bld{D}_w-\bld{W})$ is strictly positive definite, and this concludes the proof.

Corollary \ref{2cor:1} is important because its states that a limiting condition of the proposed model in \eqref{2eq:condY1} makes it a traditional CAR model, and hence the traditional CAR model is a special case of this TAR\textsubscript{C} model. However, the distribution of $\bld{y}$ for the traditional CAR model is not proper since $\bld{D}_w-\bld{W}$ is not strictly positive definite (because $(\bld{D}_{w}-\bld{W})\bld{1}=\bld{0}_{n}$). To overcome this issue, $\bld{D}_w-\bld{W}$ is rewritten as $\bld{D}_w-\rho\bld{W}$ where additional restrictions are imposed on the value of $\rho$ \citep{banerjee2015hierarchical}. This fact shows several advantages of the TAR\textsubscript{C} model. Here, Corollary \ref{2cor:2} makes a very important contribution as it establishes the argument that the distribution in \eqref{2eq:condY1} is already proper because the respective precision matrix is strictly positive definite. Hence, it needs no ad hoc adjustments to the proximity matrix $\bld{W}$ in order to make the distribution proper that is required in the traditional CAR model. Additionally, there is no need to learn a $\rho$ parameter in this case, which is tricky \citep{wall2004close}. While the traditional CAR model is a special case of the TAR\textsubscript{C} model with infinite precision-nugget, a reverse relation exists between the TAR\textsubscript{C} and CAR strategy.

\begin{proposition} \label{2prop:3}
    TAR\textsubscript{C} model is a type of CAR model.
\end{proposition}
\textit{Proof:} To prove this, we need to show that the positive definite covariance matrix in \eqref{2eq:condY1} can be expressed as the covariance matrix of a CAR model $(\bld{I}-\bld{C})^{-1}\bld{M}$ (refer to Theorem 2 of \citealp{verhoef2018relationship}) for $n \times n$ matrices $\bld{C}$ and $\bld{M}$ such that $(\bld{I}-\bld{C})^{-1}\bld{M}$ is positive definite. Corollary \ref{2cor:2} shows that the precision matrix $\bld{Q}=\frac{1}{\tau_{Y}^{2}}\bld{D}_w + \frac{1}{\sigma_{Y}^{2}}(\bld{D}_w-\bld{W})$ is strictly positive definite, which means the respective covariance matrix is also strictly positive definite. Now $\bld{Q}^{-1}=\left[\frac{1}{\tau_{Y}^{2}}\bld{D}_w + \frac{1}{\sigma_{Y}^{2}}(\bld{D}_w-\bld{W})\right]^{-1} = \left[\left(\frac{1}{\tau_{Y}^{2}}+\frac{1}{\sigma_{Y}^{2}}\right)\bld{D}_w-\frac{1}{\sigma_{Y}^{2}}\bld{W}\right]^{-1} = \left[\left(\frac{1}{\tau_{Y}^{2}}+\frac{1}{\sigma_{Y}^{2}}\right)\bld{D}_w\left(\bld{I}-\frac{\tau_{Y}^{2}}{\tau_{Y}^{2}+\sigma_{Y}^{2}}\bld{D}_w^{-1}\bld{W}\right)\right]^{-1}=\left(\bld{I}-\frac{\tau_{Y}^{2}}{\tau_{Y}^{2}+\sigma_{Y}^{2}}\bld{A}\right)^{-1}\left[\left(\frac{1}{\tau_{Y}^{2}}+\frac{1}{\sigma_{Y}^{2}}\right)\bld{D}_w\right]^{-1}$, where recall $\bld{A}=\bld{D}_w^{-1}\bld{W}$ since $a_{ij} = w_{ij}/\sum_{j} w_{ij}$. Therefore, comparing it with $\bld{Q}^{-1}=(\bld{I}-\bld{C})^{-1}\bld{M}$ we get $\bld{C}=\frac{\tau_{Y}^{2}}{\tau_{Y}^{2}+\sigma_{Y}^{2}}\bld{A}$ and $\bld{M}=\left[\left(\frac{1}{\tau_{Y}^{2}}+\frac{1}{\sigma_{Y}^{2}}\right)\bld{D}_w\right]^{-1}$. These expressions satisfy conditions C1 -- C4 in \citet{verhoef2018relationship} (pg. 3) in the following way -- (C1) $\bld{M}$ is strictly diagonal matrix with $m_{ii}>0,\forall i$ which makes $\bld{M}$ and $\bld{M}^{-1}$ positive definite; also Corollary \ref{2cor:2} shows that $\bld{Q}$ and hence $\bld{Q}^{-1}$ is positive definite, so $\bld{Q}^{-1}=(\bld{I}-\bld{C})^{-1}\bld{M}$ ensures that $(\bld{I}-\bld{C})$ has positive eigenvalues (by Proposition 3 of \citealp{verhoef2018relationship}, pg. 17), (C2) $\bld{M}$ is strictly diagonal matrix with $m_{ii}>0,\forall i$ and $m_{ij}=0$ for $i \neq j$, since $\bld{D}_w$ is a strictly diagonal matrix with $\left(D_w\right)_{ii}=\sum_j w_{ij}$ and $\left(D_w\right)_{ij}=0$ for $i \neq j$, (C3) $c_{ii}=0,\forall i$ since $a_{ij} \neq 0$ only when $i \sim j$, and (C4) $c_{ij}=\frac{\tau_{Y}^{2}}{\tau_{Y}^{2}+\sigma_{Y}^{2}}a_{ij}=\left[\left(\frac{1}{\tau_{Y}^{2}}+\frac{1}{\sigma_{Y}^{2}}\right)(D_w)_{ii}\right]^{-1}\left(\frac{1}{\sigma_{Y}^{2}}w_{ij}\right)=m_{ii}\left(\frac{1}{\sigma_{Y}^{2}}w_{ij}\right) \implies c_{ij}/m_{ii}=\frac{1}{\sigma_{Y}^{2}}w_{ij}$; $w_{ij}=w_{ji}$ since $\bld{W}$ is symmetric, so we get $c_{ij}/m_{ii}=\frac{1}{\sigma_{Y}^{2}}w_{ji}=c_{ji}/m_{jj}$. Therefore, the model in \eqref{2eq:condY1} is a type of CAR model, and this concludes the proof.

Proposition \ref{2prop:3} strengthens the connection between the TAR\textsubscript{C} and the CAR. Through Proposition \ref{2prop:3} we are able to show that the TAR\textsubscript{C} model is a type of CAR model without any ad hoc introduction of $\rho$. One should be aware of the fact that any areal model can be written in the form of a traditional CAR model. For example, consider an areal model with strictly positive definite covariance matrix, $\bld{\Sigma}$. Theorem 2 of \citet{verhoef2018relationship} states that $\bld{\Sigma}$ can be expressed as $(\bld{I}-\bld{C})^{-1}\bld{M}$ which is the covariance matrix of a CAR model. This is because any strictly positive definite precision matrix can be decomposed into a diagonal and an off-diagonal matrix and one can compute $\bld{C}$ and $\bld{M}$ from them. One of our goals is to propose an alternative to the traditional CAR model reviewed in Section \ref{2subsec:reviewCAR}. The TAR\textsubscript{C} model in \eqref{2eq:condY1} is the alternative to that, that avoids the traditional specification of $\rho$.

\subsection{Truncation on Simultaneous Distribution} \label{2subsec:JointTrunc}
Unlike the truncation defined on each $Y_{i}$ given its neighbors in Section \ref{2subsec:ConditionalTrunc}, here we apply the truncation simultaneously on all $Y_{i},\forall i,\ldots,n$ in the following way:
\begin{align}
    f\Bigl(\bld{y}\vert \bld{\beta},\sigma_{Y}^{2},\tau_{Y}^{2},\bld{u}\Bigr) \propto &\left(\frac{1}{2\pi \tau_{Y}^{2}}\right)^{n/2} \exp\left(-\frac{1}{2\tau_{Y}^{2}}(\bld{y}-\bld{X\beta})'(\bld{y}-\bld{X\beta})\right) \nonumber \\
    &\prod_{i=1}^n \mathbbm{1}\left(- \sqrt{-2\sigma_{Y}^{2}\log u_{i}} < \widetilde{Y}_{i} - \sum_{j \in N_{i}} a_{ij}\widetilde{Y}_{j} < \sqrt{-2\sigma_{Y}^{2}\log u_{i}}\right), \label{2eq:trunc2}
\end{align}
where $\sigma_{Y}^{2}$ and $\tau_{Y}^{2}$ are variance parameters, and $a_{ij} \in [0,1]$ are weights defined in Section \ref{2subsec:ConditionalTrunc}. Here, through the truncation we enforce $\widetilde{Y}_{i},\forall i=1,\ldots,n$ to be jointly close to the local averages of their respective neighborhoods. The definition of truncation in \eqref{2eq:trunc2} model is different from the TAR\textsubscript{C} model since here the truncation is applied simultaneously/jointly on $\widetilde{\bld{y}}=\left(\widetilde{Y}_{1},\ldots,\widetilde{Y}_{n}\right)'$. We call the model in \eqref{2eq:trunc2} a ``\textit{simultaneous truncated autoregressive model}'' (TAR\textsubscript{S}). Again, we would want the values of $u_{i},\forall i=1,\ldots,n$ to be close to 1 so that the bounds on $\widetilde{Y}_{i},\forall i=1,\ldots,n$ are small. Similar to Section \ref{2subsec:ConditionalTrunc}, one can again interpret $\{u_{i}\}$ as auxiliary, i.e., to be integrated across variables.

\begin{proposition} \label{2prop:4}
    Upon integrating out $\bld{u}=(u_{1},\ldots,u_{n})'$, the model in \eqref{2eq:trunc2} becomes
    \begin{equation} \label{2eq:condY2}
        \bld{y}\vert \bld{\beta},\sigma_{Y}^{2},\tau_{Y}^{2} \sim \mathcal{N}\left(\bld{X\beta},\left[\frac{1}{\tau_{Y}^{2}}\bld{I} + \frac{1}{\sigma_{Y}^{2}}(\bld{I}-\bld{A})'(\bld{I}-\bld{A})\right]^{-1}\right),
    \end{equation}
    where $\bld{A}$ is row-normalized proximity matrix.
\end{proposition}
\textit{Proof:} See Appendix \ref{2app:a2}.

Notice that just like the model in \eqref{2eq:condY1}, the precision-nugget is again added in the precision matrix of $\bld{y}\vert \bld{\beta},\sigma_{Y}^{2},\tau_{Y}^{2}$ in \eqref{2eq:condY2}. This immediately leads to the following two corollaries from Proposition \ref{2prop:4}.

\begin{corollary} \label{2cor:3}
    When $\tau_{Y}^{2} = \infty$, the model in \eqref{2eq:condY2} becomes a traditional SAR model with covariance matrix $\sigma_{Y}^{2}[(\bld{I}-\bld{A})'(\bld{I}-\bld{A})]^{-1}$.
\end{corollary}
\textit{Proof:} If we set $\tau_{Y}^{2} = \infty$ into \eqref{2eq:condY2}, it becomes $\bld{y}\vert \bld{\beta},\sigma_{Y}^{2} \sim \mathcal{N}\left(\bld{X\beta},\sigma_{Y}^{2}[(\bld{I}-\bld{A})'(\bld{I}-\bld{A})]^{-1}\right)$. This is a SAR model with mean $\bld{X\beta}$ and covariance matrix $\sigma_{Y}^{2}[(\bld{I}-\bld{A})'(\bld{I}-\bld{A})]^{-1}$ (refer to Section \ref{2subsec:reviewSAR}; also see \citealp{cressie1993statistics,banerjee2015hierarchical} for standard references).

\begin{corollary} \label{2cor:4}
    The precision matrix $\frac{1}{\tau_{Y}^{2}}\bld{I} + \frac{1}{\sigma_{Y}^{2}}(\bld{I}-\bld{A})'(\bld{I}-\bld{A})$ of the distribution in \eqref{2eq:condY2} is strictly positive definite.
\end{corollary}
\textit{Proof:} Here also we calculate $\bld{v}'\left[\frac{1}{\tau_{Y}^{2}}\bld{I} + \frac{1}{\sigma_{Y}^{2}}(\bld{I}-\bld{A})'(\bld{I}-\bld{A})\right]\bld{v} = \frac{1}{\tau_{Y}^{2}}\bld{v}'\bld{v} + \frac{1}{\sigma_{Y}^{2}}\bld{v}'(\bld{I}-\bld{A})'(\bld{I}-\bld{A})\bld{v}$ for any non-zero vector $\bld{v}$. We get $\bld{v}'(\bld{I}-\bld{A})'(\bld{I}-\bld{A})\bld{v} = [(\bld{I}-\bld{A})\bld{v}]'[(\bld{I}-\bld{A})\bld{v}] \geq 0$ for any non-zero vector $\bld{v}$. Also, $\bld{v}'\bld{v} = \sum_{i}v_{i}^{2} > 0$. Therefore, $\bld{v}'\left[\frac{1}{\tau_{Y}^{2}}\bld{I} + \frac{1}{\sigma_{Y}^{2}}(\bld{I}-\bld{A})'(\bld{I}-\bld{A})\right]\bld{v} > 0$. Hence, $\frac{1}{\tau_{Y}^{2}}\bld{I} + \frac{1}{\sigma_{Y}^{2}}(\bld{I}-\bld{A})'(\bld{I}-\bld{A})$ is strictly positive definite, and this concludes the proof.

Corollary \ref{2cor:3} shows that a limiting condition of the model in \eqref{2eq:condY2} makes it a SAR model, which means the proposed model is comparable to a SAR model. However, like a CAR model, the distribution of $\bld{y}$ for a traditional SAR model is also improper since $(\bld{I}-\bld{A})'(\bld{I}-\bld{A})$ is not strictly positive definite although it is symmetric. Therefore, adjustments are needed in the precision matrix and a typical choice for $\bld{A}$ is $\rho\bld{A}$, where $\rho$ is called a spatial autocorrelation parameter. Moreover, $(\bld{I}-\rho\bld{A})'(\bld{I}-\rho\bld{A})$ is strictly positive definite only when $\rho \in (-1,1)$ \citep{banerjee2015hierarchical}, which means additional ad hoc adjustments are imposed on $\rho$. Corollary \ref{2cor:4} is particularly important here because it shows that the distribution in \eqref{2eq:condY2} is already proper because the respective precision matrix is strictly positive definite. Hence, no ad hoc adjustments to $\bld{A}$ is needed which further means no need to learn a $\rho$ parameter which is very difficult. Therefore, the model in $\eqref{2eq:condY2}$ can be seen as advantageous from this perspective. Similar to the SAR/CAR and TAR\textsubscript{C}/CAR relationships, we have the following results

\begin{proposition} \label{2prop:5}
    TAR\textsubscript{S} model can be uniquely written as a version of the TAR\textsubscript{C} model.
\end{proposition}
\textit{Proof:} To prove this, we need to show that the covariance matrix of $\bld{y}\vert \bld{\beta},\sigma_{Y}^{2},\tau_{Y}^{2}$ in \eqref{2eq:condY2} can be written as a version of the covariance matrix of $\bld{y}\vert \bld{\beta},\sigma_{Y}^{2},\tau_{Y}^{2}$ in \eqref{2eq:condY1}. From the proof of Proposition \ref{2prop:3}, we see that the covariance matrix of $\bld{y}\vert \bld{\beta},\sigma_{Y}^{2},\tau_{Y}^{2}$ in \eqref{2eq:condY1} can be written as $\bld{Q}^{-1}=(\bld{I}-\bld{C})^{-1}\bld{M}=\left(\bld{I}-\frac{\tau_{Y}^{2}}{\tau_{Y}^{2}+\sigma_{Y}^{2}}\bld{A}\right)^{-1}\left[\left(\frac{1}{\tau_{Y}^{2}}+\frac{1}{\sigma_{Y}^{2}}\right)\bld{D}_w\right]^{-1}$. Now, we try to write the covariance matrix of $\bld{y}\vert \bld{\beta},\sigma_{Y}^{2},\tau_{Y}^{2}$ in \eqref{2eq:condY2} in terms of $(\bld{I}-\bld{C})^{-1}\bld{M}$. To do so, we expand the respective precision matrix $\bld{Q}$ as $\bld{Q}=\frac{1}{\tau_{Y}^{2}}\bld{I} + \frac{1}{\sigma_{Y}^{2}}(\bld{I}-\bld{A})'(\bld{I}-\bld{A})=\left[\left(\frac{1}{\tau_{Y}^{2}}+\frac{1}{\sigma_{Y}^{2}}\right)\bld{I}+\frac{1}{\sigma_{Y}^{2}}\bld{D}_1\right]-\frac{1}{\sigma_{Y}^{2}}\left(\bld{A}+\bld{A}'-\bld{D}_2\right)$, where we write $\bld{A}'\bld{A}=\bld{D}_1 + \bld{D}_2$. Here, $\bld{D}_1$ is a diagonal matrix with $\mathrm{diag}(\bld{D}_1)=\mathrm{diag}(\bld{A}'\bld{A})$, and $\bld{D}_2$ is a matrix with $\mathrm{diag}(\bld{D}_2)=\bld{0}$ and off-diagonal elements of $\bld{D}_2$ are same as the off-diagonal elements of matrix $\bld{A}'\bld{A}$. Comparing the aforementioned expression of $\bld{Q}$ with $\bld{Q}=\bld{D}-\bld{R}$ (refer to Item (i) of Theorem 2 of \citealp{verhoef2018relationship}, pg. 5) we get $\bld{D}=\left(\frac{1}{\tau_{Y}^{2}}+\frac{1}{\sigma_{Y}^{2}}\right)\bld{I}+\frac{1}{\sigma_{Y}^{2}}\bld{D}_1$, and $\bld{R}=\frac{1}{\sigma_{Y}^{2}}\left(\bld{A}+\bld{A}'-\bld{D}_2\right)$. Following Item (ii) of Theorem 2 in \citealp{verhoef2018relationship} (pg. 5), we get $\bld{C}=\left[\left(\frac{1}{\tau_{Y}^{2}}+\frac{1}{\sigma_{Y}^{2}}\right)\bld{I}+\frac{1}{\sigma_{Y}^{2}}\bld{D}_1\right]^{-1}\frac{1}{\sigma_{Y}^{2}}\left(\bld{A}+\bld{A}'-\bld{D}_2\right)$ and $\bld{M}=\left[\left(\frac{1}{\tau_{Y}^{2}}+\frac{1}{\sigma_{Y}^{2}}\right)\bld{I}+\frac{1}{\sigma_{Y}^{2}}\bld{D}_1\right]^{-1}$, where $\bld{Q}^{-1}=(\bld{I}-\bld{C})^{-1}\bld{M}$. Hence, according to Theorem 2 in \citet{verhoef2018relationship} (pg. 5), $\bld{C}$ and $\bld{M}$ are uniquely determined by $\bld{Q}^{-1}$, and this concludes the proof.

\begin{corollary} \label{2cor:5}
    TAR\textsubscript{S} model is a type of CAR model.
\end{corollary}
\textit{Proof:} It follows immediately from Propositions \ref{2prop:3} and \ref{2prop:5}.

We see the results in Corollaries \ref{2cor:1} and \ref{2cor:3} as a way to see both TAR\textsubscript{C} and TAR\textsubscript{S} as a more flexible modeling framework as the CAR and SAR models are special cases when the precision-nugget $\tau_{Y}^{2}=\infty$, which may actually be finite. At the same time we see the reverse relationship (when CAR/SAR is equivalent to TAR) in Proposition \ref{2prop:3} and Corollary \ref{2cor:5} as an advantage as the CAR/SAR are well known to be sensible models.

The SAR model is closely related to the widely known nearest-neighbor Gaussian process \citep[NNGP;][]{datta2016nngp,datta2022nearest} model in point-referenced spatial data setting through the covariance matrix structure. Naturally, a connection can be established between the TAR and NNGP model. We consider a different local average to define TAR\textsubscript{S} in the point-referenced data setting, which leads to the NNGP as a special case. A detailed discussion on this is provided in Appendix \ref{2app:b}. This special case is particularly interesting as it shows a way for someone to modify a TAR model for their setting. Specifically, one might consider different priors on $\{u_{i}\}$ and can replace the local average $\sum_{j \in N_{i}} a_{ij}\widetilde{Y}_{j}$ within the truncation with their favorite local average (in Appendix \ref{2app:b} $\sum_{j \in N_{i}} a_{ij}\widetilde{Y}_{j}$ is replaced with a specific inverse distance weighting predictor).

\subsection{Bayesian Implementation} \label{2subsec:BHM}
We consider the case when some data may be missing. That is, specify the last $n_{\mathcal{M}}$ regions of $\{1,\ldots,n\}$ to be missing. So, the first $(n-n_{\mathcal{M}})=n_{O}$ regions are observed, where we note that the ordering of regions $\{1,\ldots,n\}$ is arbitrary. We define $\bld{O}=\begin{pmatrix} \bld{I}_{n_{O}} & \bld{0}_{n_{O} \times n_{\mathcal{M}}} \end{pmatrix}$ to be an $n_{O} \times n$ dimensional incidence matrix. Notice that, multiplying $\bld{O}$ with an $n$-dimensional vector (or an $n \times p$ dimensional matrix) gives us the vector (or matrix) at the observed locations. So, $\bld{Oy}=(Y_{1},\ldots,Y_{n_{O}})'=\bld{y}_{O}$ and $\bld{OX}=\left(\bld{x}_{1},\ldots,\bld{x}_{n_{O}}\right)'=\bld{X}_{O}$. We write a general form $\bld{y}\vert \bld{\beta},\sigma_{Y}^{2},\delta \sim \mathcal{N}\left(\bld{X\beta},\sigma_{Y}^{2}\bld{\Sigma}(\delta)\right)$ that represents the Equations \eqref{2eq:condY1} and \eqref{2eq:condY2} where we set $\tau_{Y}^{2}=\delta\sigma_{Y}^{2}$ with $\delta > 0$. Therefore, we define $\bld{\Sigma}(\delta)=\left[(1/\delta)\bld{D}_{w}+(\bld{D}_{w}-\bld{W})\right]^{-1}$ for TAR\textsubscript{C} model in \eqref{2eq:condY1}, and $\bld{\Sigma}(\delta)=\left[(1/\delta)\bld{I}+(\bld{I}-\bld{A})'(\bld{I}-\bld{A})\right]^{-1}$ for TAR\textsubscript{S} model in \eqref{2eq:condY2}.

We now write a generic Bayesian hierarchical model where the resulting joint distribution is proportional to the product of the following conditional and marginal distributions:
\begin{align}
    \mathrm{Data\hspace{5pt}Model:}\hspace{5pt}& \bld{y}_{O}\vert \bld{\beta},\sigma_{Y}^{2},\delta \sim \mathcal{N}\left(\bld{X}_{O}\bld{\beta},\sigma_{Y}^{2}\left(\bld{O}\bld{\Sigma}(\delta)^{-1}\bld{O}'\right)^{-1}\right), \nonumber \\
    \mathrm{Parameter\hspace{5pt}Model\hspace{5pt}1:}\hspace{5pt}& \pi(\bld{\beta}) = 1, \nonumber \\
    \mathrm{Parameter\hspace{5pt}Model\hspace{5pt}2:}\hspace{5pt}& \sigma_{Y}^{2} \sim \mathcal{IG}(a,b), \nonumber \\
    \mathrm{Parameter\hspace{5pt}Model\hspace{5pt}3:}\hspace{5pt}& \delta \sim \mathcal{DU}(\delta_{1},\ldots,\delta_{k}), \label{2eq:BHM}
\end{align}
where $\bld{O}\bld{\Sigma}(\delta)^{-1}\bld{O}'$ gives us the precision matrix of $\bld{y}$ at the observed locations, $\bld{\beta}$ is assigned a flat prior, $\sigma_{Y}^{2}$ is assigned an inverse gamma prior with shape and scale parameter as $a$ and $b$ respectively, and $\delta$ is assigned a discrete uniform prior with support $\{\delta_{1},\ldots,\delta_{k}\}$. Alternative strategies to placing a discrete uniform distribution on $\delta$ include other model combination strategies such as predictive stacking \citep{zhang2023exact} or to simply choose a value of $\{\delta_{1},\ldots, \delta_{k}\}$ using an information criterion. Here, we generate samples directly from the posterior distribution $f(\bld{\beta},\sigma_{Y}^{2},\delta\vert \bld{y}_{O})$. The advantage of assigning a flat prior to $\bld{\beta}$ is that it ensures conjugacy for $\sigma_{Y}^{2}$. \citet{banerjee2020conjugate} used a normal prior for $\bld{\beta}$ to attain conjugacy. But in order to do that, they had to adjust $\sigma_{Y}^{2}$ in the prior for $\bld{\beta}$. The hierarchical model in \eqref{2eq:BHM} does not need adjustment for $\sigma_{Y}^{2}$ in $\pi(\bld{\beta})$ to attain conjugacy, and we achieve that by assigning a flat prior to $\bld{\beta}$.

Suppose, we wish to sample from the joint distribution of random variables say $X$ and $Y$. To do so, one can use a ``composite sampler'' \citep{varin2011overview,saha2024subsampling} where we first sample $Y$ from its density $f(Y)$ and then sample $X$ from the conditional density $f(X\vert Y)$. We derive the posterior distributions of $Y$ and $X\vert Y$ through conjugacy, and sample directly from these distributions without needing an MCMC method. Recently, samplers of this type have been used in many papers \citep{zhang2019practical,banerjee2020conjugate,bradley2023deep,bradley2024generating}. The benefit of using such sampler is that it is much faster since it avoids sampling through MCMC which usually takes time to reach the stationary distribution. For the hierarchical model in \eqref{2eq:BHM} our goal is to sample from $f(\bld{\beta},\sigma_{Y}^{2},\delta\vert\bld{y}_{O})=f(\bld{\beta}\vert\sigma_{Y}^{2},\delta,\bld{y}_{O})f(\sigma_{Y}^{2}\vert\delta,\bld{y}_{O})f(\delta\vert\bld{y}_{O})$, hence we first sample from $f(\delta\vert\bld{y}_{O})$, then sample from $f(\sigma_{Y}^{2}\vert\delta,\bld{y}_{O})$ for each sample of $\sigma_{Y}^{2}$, and finally sample from $f(\bld{\beta}\vert\sigma_{Y}^{2},\delta,\bld{y}_{O})$ for each sample of $(\sigma_{Y}^{2},\delta)$. These distributions are given by
\begin{align}
    \bld{\beta}\vert\sigma_{Y}^{2},\delta,\bld{y}_{O} &\sim \mathcal{N}\Bigl((\bld{X}_{O}'\bld{\Sigma}_{o}(\delta)^{-1}\bld{X}_{O})^{-1}\bld{X}_{O}'\bld{\Sigma}_{o}(\delta)^{-1}\bld{y}_{O},\sigma_{Y}^{2}(\bld{X}_{O}'\bld{\Sigma}_{o}(\delta)^{-1}\bld{X}_{O})^{-1}\Bigr), \nonumber \\
    \sigma_{Y}^{2}\vert\delta,\bld{y}_{O} &\sim \mathcal{IG}\Bigl(a+\frac{n_{O}-p}{2}, \nonumber \\
    &\qquad b+\frac{1}{2}\left[\bld{y}_{O}'\bld{\Sigma}_{o}(\delta)^{-1}\bld{y}_{O} - \bld{y}_{O}'\bld{\Sigma}_{o}(\delta)^{-1}\bld{X}_{O}(\bld{X}_{O}'\bld{\Sigma}_{o}(\delta)^{-1}\bld{X}_{O})^{-1}\bld{X}_{O}'\bld{\Sigma}_{o}(\delta)^{-1}\bld{y}_{O}\right]\Bigr), \nonumber \\
    f(\delta\vert\bld{y}_{O}) &= \frac{f(\delta,\bld{y}_{O})}{\sum_{\delta \in \{\delta_{1},\ldots,\delta_{k}\}}f(\delta,\bld{y}_{O})}, \label{2eq:postdist}
\end{align}
where $\bld{\Sigma}_{o}(\delta)^{-1}=\bld{O}\bld{\Sigma}(\delta)^{-1}\bld{O}'$, and the expression of $f(\delta,\bld{y}_{O})$ as well as the proof of \eqref{2eq:postdist} is given in Appendix \ref{2app:a3}. Notice that since Equation \eqref{2eq:postdist} is derived with respect to $\bld{\Sigma}_{o}(\delta)^{-1}$, no inverse operation of an $n_{O} \times n_{O}$ dimensional matrix is needed, and hence sampling from \eqref{2eq:postdist} is very fast and the model can be extended to very large datasets. 

\subsection{Posterior Predictive Inference} \label{2subsec:postpred}
We obtain the posterior predictive distribution at the missing regions/locations using the standard Kriging predictor. We rewrite the generic joint distribution of the data at the observed and missing regions/locations in the following way:
\begin{equation} \label{2eq:jointpred}
    \begin{pmatrix}
        \bld{y}_{O}\\
        \bld{y}_{\mathcal{M}}
    \end{pmatrix} \vert \hspace{0.2cm} \bld{\beta},\sigma_{Y}^{2},\delta
    \sim \mathcal{N}\Biggl(
    \begin{pmatrix}
        \bld{X}_{O}\bld{\beta}\\
        \bld{X}_{\mathcal{M}}\bld{\beta}
    \end{pmatrix},
    \sigma_{Y}^{2}
    \begin{pmatrix}
        \bld{\Sigma}_{n_{O},n_{O}}(\delta) & \bld{\Sigma}_{n_{O},n_{\mathcal{M}}}(\delta)\\
        \bld{\Sigma}_{n_{\mathcal{M}},n_{O}}(\delta) & \bld{\Sigma}_{n_{\mathcal{M}},n_{\mathcal{M}}}(\delta)
    \end{pmatrix}
    \Biggr),
\end{equation}
where we write the $n \times n$ covariance matrix $\bld{\Sigma}(\delta)$ as $\begin{pmatrix} \bld{\Sigma}_{n_{O},n_{O}}(\delta) & \bld{\Sigma}_{n_{O},n_{\mathcal{M}}}(\delta) \\ \bld{\Sigma}_{n_{\mathcal{M}},n_{O}}(\delta) & \bld{\Sigma}_{n_{\mathcal{M}},n_{\mathcal{M}}}(\delta) \end{pmatrix}$. We reiterate that $\bld{\Sigma}(\delta)=\bigl[(1/\delta)\bld{D}_{w}+(\bld{D}_{w}-\bld{W})\bigr]^{-1}$ for TAR\textsubscript{C} model, and $\bld{\Sigma}(\delta)=\bigl[(1/\delta)\bld{I}+(\bld{I}-\bld{A})'(\bld{I}-\bld{A})\bigr]^{-1}$ for TAR\textsubscript{S} model. Through standard distribution theory \citep{ravishanker2021first} we get
\begin{align}
    \bld{y}_{\mathcal{M}}\vert\bld{y}_{O},\bld{\beta},\sigma_{Y}^{2},\delta \sim \mathcal{N}\Bigl(&\bld{X}_{\mathcal{M}}\bld{\beta}+\bld{\Sigma}_{n_{\mathcal{M}},n_{O}}(\delta)\bld{\Sigma}_{n_{O},n_{O}}(\delta)^{-1}(\bld{y}_{O}-\bld{X}_{O}\bld{\beta}), \nonumber \\
    &\qquad \sigma_{Y}^{2}\left\{\bld{\Sigma}_{n_{\mathcal{M}},n_{\mathcal{M}}}(\delta)-\bld{\Sigma}_{n_{\mathcal{M}},n_{O}}(\delta)\bld{\Sigma}_{n_{O},n_{O}}(\delta)^{-1}\bld{\Sigma}_{n_{O},n_{\mathcal{M}}}(\delta)\right\}\Bigr). \label{2eq:postpred}
\end{align}
Sampling from the posterior predictive distribution in \eqref{2eq:postpred} is computationally very efficient because (i) $\bld{\Sigma}(\delta)$ needs to be computed only \textit{once} (from known $\bld{\Sigma}(\delta)^{-1}$) for $\delta=1$, (ii) one can sample the $i$-th element of $\bld{y}_{\mathcal{M}}$ in parallel rather than a joint distribution for a set of missing regions/locations which is incredibly faster and (iii) because of the expressions of $\bld{\Sigma}(\delta)^{-1}$ one can efficiently apply Neumann series \citep{ravishanker2021first} to find $\bld{\Sigma}(\delta)$ from $\bld{\Sigma}(\delta)^{-1}$ in case of very big dataset. The Neumann series is an inverse identity that states, $(\lambda\bld{I}+\bld{T})^{-1}=\sum_{k=0}^{\infty}(-1)^{k}(1/\lambda)^{k+1}\bld{T}^{k}$, where $\lambda$ is a constant, $\bld{I}$ is an identity matrix and $\bld{T}$ is a square matrix. We calculate the point prediction by taking the mean of the posterior samples drawn for $\bld{y}_{\mathcal{M}}$.

\section{Simulations} \label{2sec:simulation}
In this section, we discuss the inferential, predictive and computational performance of the proposed TAR model through several illustrations. Our goal through the simulation study is to demonstrate the model's performance on areal datasets under several simulation settings (Section \ref{2subsec:illusCarSar}) and compare it with the traditional CAR and SAR models while considering multiple replicates (Section \ref{2subsec:simstudyCarSar}). We also compare the computation time among the TAR, CAR and SAR model (Section \ref{2subsec:comptime}). Specifically, we consider the scenarios when the data is distributed according to CAR, SAR, TAR\textsubscript{C} and TAR\textsubscript{S}, and compare inferential and predictive performance of all models in each simulation setting.

\begin{figure}[t!]
    \centering
    \includegraphics[scale=1]{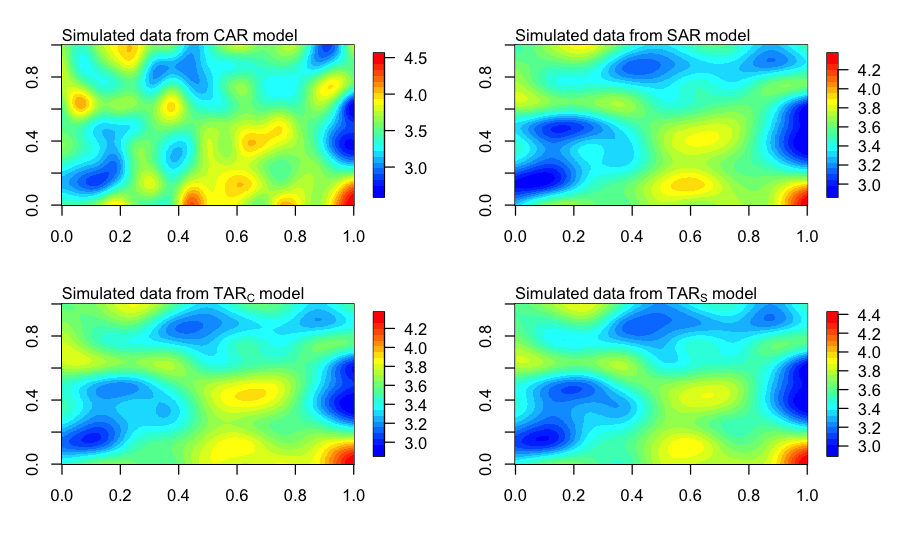}
    \caption{Image plot of the simulated data based on the simulation settings mentioned in Items (a) -- (d) with $\bld{\beta}=(2,5)'$, $\sigma_{Y}^{2}=0.5$, $\delta=1$ and $\rho=-0.606$.}
    \label{2fig:simdata}
\end{figure}

\begin{figure}[t!]
    \centering
    \includegraphics[scale=0.7]{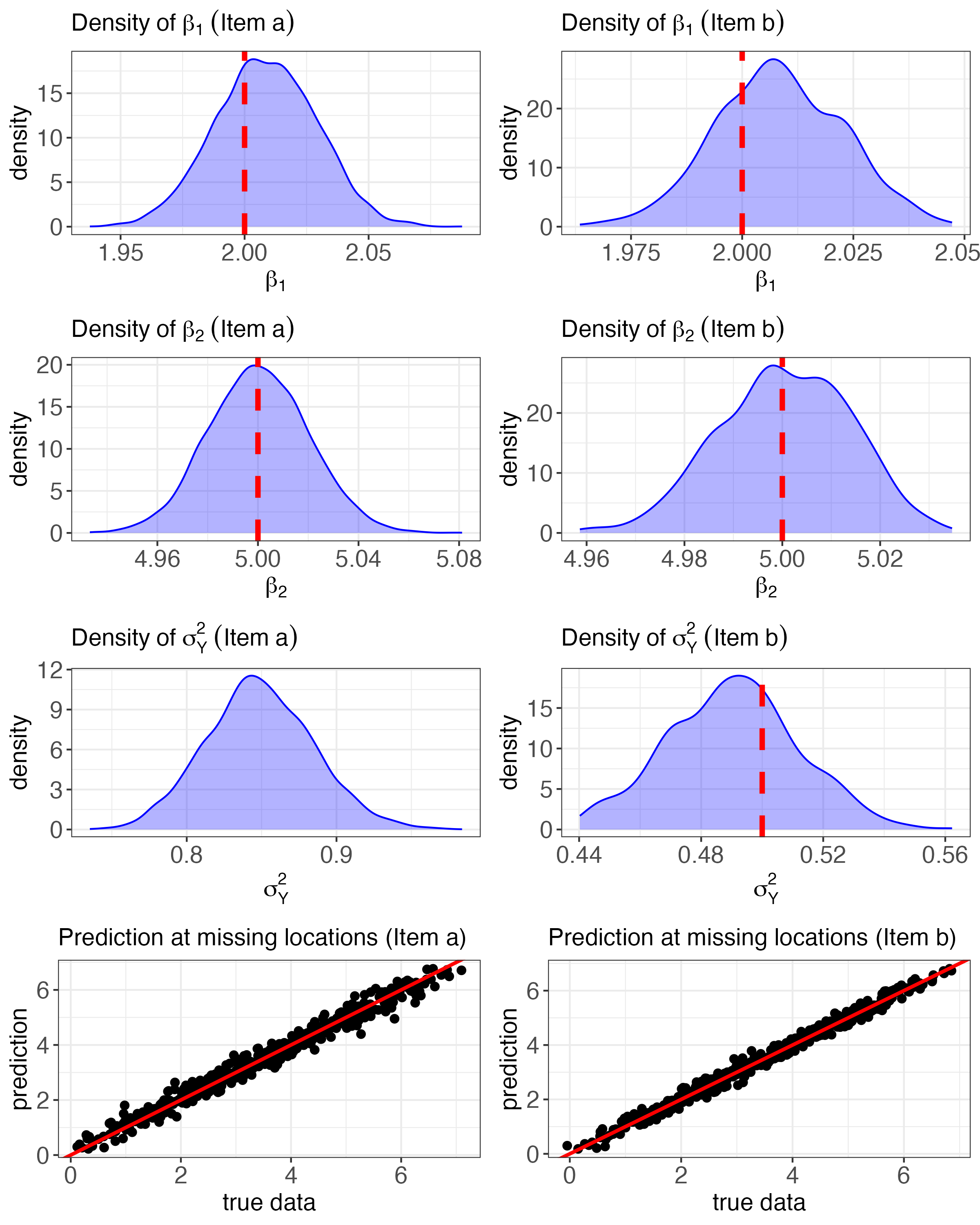}
    \caption{We plot the density of posterior samples for $\beta_{1}$ (first row), $\beta_{2}$ (second row) and $\sigma_{Y}^{2}$ (third row) when implementing the TAR\textsubscript{C} model with $\delta=1$. The vertical dashed lines represent the true values $\beta_{1}=2$, $\beta_{2}=5$ and $\sigma_{Y}^{2}=0.5$. We also plot the predictions at the missing locations vs. the true data (fourth row), where the straight line represents the 45{\degree} reference line. The plots on the left and right column are generated using the simulated datasets based on Items (a) and (b) respectively.}
    \label{2fig:simTCAR}
\end{figure}

\subsection{Illustration on Areal Data} \label{2subsec:illusCarSar}
We simulate $n=1600$ spatial locations on a $40 \times 40$ grid over $[0,1] \times [0,1]$. We create the proximity matrix $\bld{W}$ based on the first-order neighborhood system, where the first-order neighbors of a certain location are the immediate adjacent locations on the grid structure. We simulate $\bld{X}$, which is a $1600 \times 2$ dimensional matrix, from an independent $\mathcal{U}(0,1)$ distribution. For $\bld{\beta}=\left(\beta_{1},\beta_{2}\right)'=(2,5)'$, $\sigma_{Y}^{2}=0.5$, $\delta=1$ and $\rho=-0.606$, $\bld{y}$ is simulated from a multivariate normal distribution with mean vector $\bld{X\beta}$ and covariance matrix $\bld{\Sigma}_{Y}$. Here, we consider four settings of $\bld{\Sigma}_{Y}$ --
\begin{enumerate}[(a)]
    \item $\bld{\Sigma}_{Y}=\sigma_{Y}^{2}\left(\bld{D}_{w}-\rho\bld{W}\right)^{-1}$ to generate $\bld{y}$ from a traditional CAR model,
    \item $\bld{\Sigma}_{Y}=\sigma_{Y}^{2}\left[(1/\delta)\bld{D}_{w} + (\bld{D}_{w}-\bld{W})\right]^{-1}$ to generate $\bld{y}$ from the TAR\textsubscript{C} model,
    \item $\bld{\Sigma}_{Y}=\sigma_{Y}^{2}\left[(\bld{I}-\rho\bld{A})'(\bld{I}-\rho\bld{A})\right]^{-1}$ to generate $\bld{y}$ from a traditional SAR model,
    \item $\bld{\Sigma}_{Y}=\sigma_{Y}^{2}\left[(1/\delta)\bld{I} + (\bld{I}-\bld{A})'(\bld{I}-\bld{A})\right]^{-1}$ to generate $\bld{y}$ from the TAR\textsubscript{S} model.
\end{enumerate}
Thus, there are times when TAR is misspecified and CAR/SAR are correctly specified and vice versa. This simulation design is chosen to demonstrate that the TAR model produces reasonable inferences even when CAR/SAR is the correct model. In Figure \ref{2fig:simdata}, we show image plots of the simulated spatial datasets based on the four aforementioned simulation settings. We enforce 16\% missing values at random which represent random areas to be missing on a geographical map. 14\% more values are enforced to be missing on an arbitrary $15 \times 15$ grid (total number of missing values = 480) that represent a large block of multiple areas to be missing on a map. We aim to make predictions on these missing locations/areas. For simplicity, we set $\delta=1$ as known when fitting TAR\textsubscript{C} and TAR\textsubscript{S} models so that $\tau_{Y}^{2} = \sigma_{Y}^{2}$.

\begin{table}[t!]
    \renewcommand{\arraystretch}{1.1}
    \centering
    \begin{tabular}{c@{\hspace{0.3in}}c@{\hspace{0.4in}}c@{\hspace{0.4in}}c}
        \hline
        \hline
        \multirow{2}{*}{} & \multirow{2}{*}{\textbf{True}} & \multicolumn{2}{c}{\centering \textbf{Simulation Setting}}\\
        & & \textbf{Item (a)} & \textbf{Item (b)}\\
        \hline
        $\beta_1$ & 2 & 2.0086 (1.97, 2.05) & 2.0083 (1.98, 2.04)\\
        $\beta_2$ & 5 & 5.004 (4.96, 5.04) & 5.0005 (4.97, 5.02)\\
        $\sigma_{Y}^{2}$ & 0.5 & 0.85 (0.78, 0.92) & 0.49 (0.45, 0.53)\\
        $R^{2}$ & - & 0.977 & 0.990\\
        MAE & - & 0.183 & 0.128\\
        RMSE & - & 0.238 & 0.160\\
        CRPS & - & 0.139 & 0.090\\
        INT & - & 1.676 & 0.762\\
        CVG & - & 0.875 & 0.940\\
        Model Runtime (sec.) & - & 0.013 & 0.014\\
        Prediction time (sec.) & - & 0.36 & 0.34\\
        \hline
        \hline
    \end{tabular}
    \caption{Parameter estimates with 95\% credible intervals, along with prediction accuracy measure and CPU time when implementing the TAR\textsubscript{C} model.}
    \label{2table:simTCAR}
\end{table}

\begin{figure}[t!]
    \centering
    \includegraphics[scale=0.7]{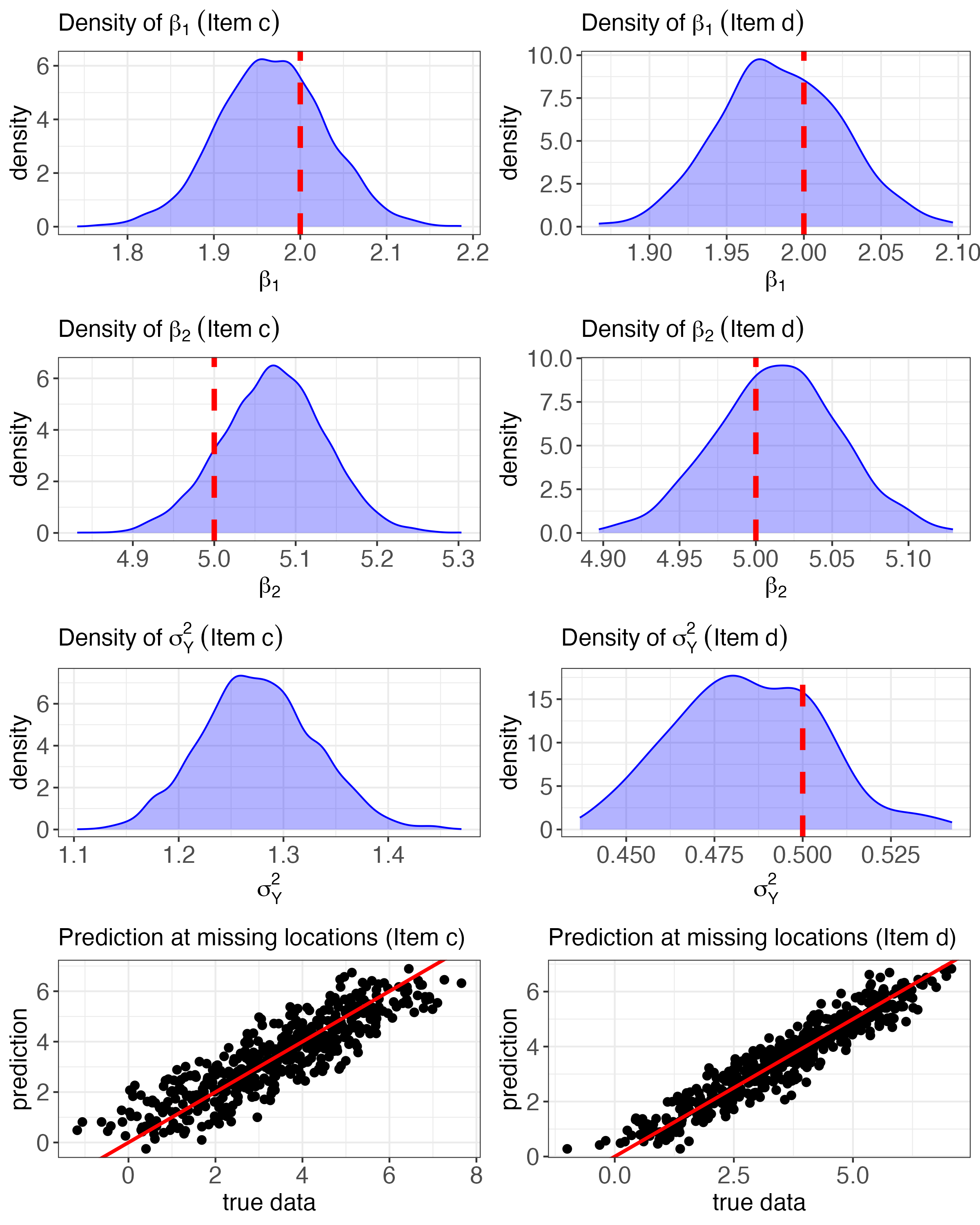}
    \caption{We plot the density of posterior samples for $\beta_{1}$ (first row), $\beta_{2}$ (second row) and $\sigma_{Y}^{2}$ (third row) when implementing the TAR\textsubscript{S} model with $\delta=1$. The vertical dashed lines represent the true values $\beta_{1}=2$, $\beta_{2}=5$ and $\sigma_{Y}^{2}=0.5$. We also plot the predictions at the missing locations vs. the true data (fourth row), where the straight line represents the 45{\degree} reference line. The plots on the left and right column are generated using the simulated datasets based on Items (c) and (d) respectively.}
    \label{2fig:simTSAR}
\end{figure}

First, we implement the TAR\textsubscript{C} model on the simulated datasets based on Items (a) and (b). We generate $G=500$ posterior samples of $\sigma_{Y}^{2}$ and $\bld{\beta}$ using Equation \eqref{2eq:postdist}. We estimate the parameters using the observed data of size 1120, and make predictions at the 480 missing locations. For each of the missing locations, we sample from the posterior predictive distribution in \eqref{2eq:postpred}. The point prediction at each location is the mean of all the $G=500$ posterior predictive samples. In Figure \ref{2fig:simTCAR}, we plot the density of the posterior samples of $\bld{\beta}$ and $\sigma_{Y}^{2}$, and the point predictions vs. the true data at the missing locations, using both the simulated datasets from Items (a) and (b). For both the simulation settings, we see that the posterior samples of $\bld{\beta}$ are distributed about the respective true values. But, $\sigma_{Y}^{2}$ is closer to the true value when the data is simulated from Item (b), otherwise it is far from the true value. From both the prediction plots we see that the predicted values at the missing locations are very close to the respective true values as they appear along the 45{\degree} reference line. We also notice that the prediction accuracy is higher for the simulated setting in Item (b). This is not unusual because Item (b) generates the data from the TAR\textsubscript{C} model itself. To draw a more precise picture, in Table \ref{2table:simTCAR} we provide the point estimates of $\left(\bld{\beta},\sigma_{Y}^{2}\right)$ with 95\% credible intervals, $R^{2}$ value, mean absolute error (MAE), root mean squared error (RMSE), continuous rank probability score (CRPS; see \citealp{gneiting2005calibrated}), interval score (INT; see \citealp{gneiting2007strictly}) and prediction interval coverage score (CVG; see \citealp{heaton2019case}) at $\alpha=0.05$ between the true and predicted values at missing locations. We also provide the model runtime and prediction time for both simulation settings. It is clear from the table that the point estimates of $\bld{\beta}$ and $\sigma_{Y}^{2}$ are closer to the true value with less variability when the data is simulated from the model itself. Furthermore, all the metrics for prediction accuracy show that the predictions at missing locations are more accurate when the data is simulated from the TAR\textsubscript{C} model itself. In both cases, notice that the model runs very fast (takes only 13 milliseconds) because we sample the posterior directly. It also does a very fast prediction and only takes about 0.36 seconds (in both cases) to predict $\bld{y}$ at all 480 missing locations. Although TAR\textsubscript{C} model produces better results when the data is generated from the model itself, it does a very good job even when the data is generated from a traditional CAR model. This shows that TAR\textsubscript{C} model is capable of capturing the spatial dependence properly.

\begin{table}[t!]
    \renewcommand{\arraystretch}{1.1}
    \centering
    \begin{tabular}{c@{\hspace{0.3in}}c@{\hspace{0.4in}}c@{\hspace{0.4in}}c}
        \hline
        \hline
        \multirow{2}{*}{} & \multirow{2}{*}{\textbf{True}} & \multicolumn{2}{c}{\centering \textbf{Simulation Setting}}\\
        & & \textbf{Item (c)} & \textbf{Item (d)}\\
        \hline
        $\beta_1$ & 2 & 1.967 (1.84, 2.09) & 1.985 (1.91, 2.06)\\
        $\beta_2$ & 5 & 5.072 (4.95, 5.19) & 5.014 (4.94, 5.09)\\
        $\sigma_{Y}^{2}$ & 0.5 & 1.276 (1.18, 1.38) & 0.48 (0.45, 0.53)\\
        $R^{2}$ & - & 0.775 & 0.906\\
        MAE & - & 0.662 & 0.385\\
        RMSE & - & 0.818 & 0.481\\
        CRPS & - & 0.490 & 0.272\\
        INT & - & 5.75 & 2.21\\
        CVG & - & 0.814 & 0.958\\
        Model Runtime (sec.) & - & 0.009 & 0.009\\
        Prediction time (sec.) & - & 0.35 & 0.34\\
        \hline
        \hline
    \end{tabular}
    \caption{Parameter estimates with 95\% credible intervals, along with prediction accuracy measure and CPU time when implementing the TAR\textsubscript{S} model.}
    \label{2table:simTSAR}
\end{table}

Now, we implement the TAR\textsubscript{S} model on the simulated datasets based on Items (c) and (d). We again generate $G=500$ posterior samples of $\sigma_{Y}^{2}$ and $\bld{\beta}$ using Equation \eqref{2eq:postdist}, and follow the same procedure as before to estimate the parameters and make predictions on the missing locations. In Figure \ref{2fig:simTSAR}, we plot the density of the posterior samples of $\bld{\beta}$, and the point predictions vs. the true data at the missing locations, using both the simulated datasets from Items (c) and (d). Notice that TAR\textsubscript{S} model does a better job in estimating $\bld{\beta}$ and $\sigma_{Y}^{2}$ when the data is simulated from the model itself. Here, they both are closer to the true value with less variability (e.g., smaller credible intervals). Furthermore, we observe from Figure \ref{2fig:simTSAR} that the predictions at the missing locations are more accurate as the scatter plot (in the last row) aligns more closely to the 45{\degree} reference line. In Table \ref{2table:simTSAR}, we provide the point estimates of $\bld{\beta}$ and $\sigma_{Y}^{2}$ with their corresponding 95\% credible intervals, $R^{2}$ value, MAE, RMSE, CRPS, INT and CVG (at $\alpha=0.05$) score of the predictions at the missing locations. Notice that, although the distribution of $\bld{\beta}$ is a little shifted from the true value when the data is simulated from a SAR model, the true value still appears within the 95\% credible interval. The interval of $\bld{\beta}$ is also smaller when the data is simulated from the model itself. We also see that the estimated $\sigma_{Y}^{2}$ is far from the true value for the simulation setting in Item (c), but it is almost same as the true value when the data is simulated from Item (d). We see a higher value in $R^{2}$ and CVG, and lower values in all the other metrics which implies a better prediction accuracy at the missing locations. Again in both cases, the model runs quickly (takes about 9 milliseconds) and also does fast prediction (takes about 0.34 seconds). All these results indicate that the TAR\textsubscript{S} model is able to capture the spatial dependence of the dataset. Although we see that the proposed model successfully identifies the spatial dependence, we are interested in comparing its performance with the traditional models (CAR and SAR) which we present in the next section.

\begin{figure}[t!]
    \centering
    \includegraphics[scale=0.63]{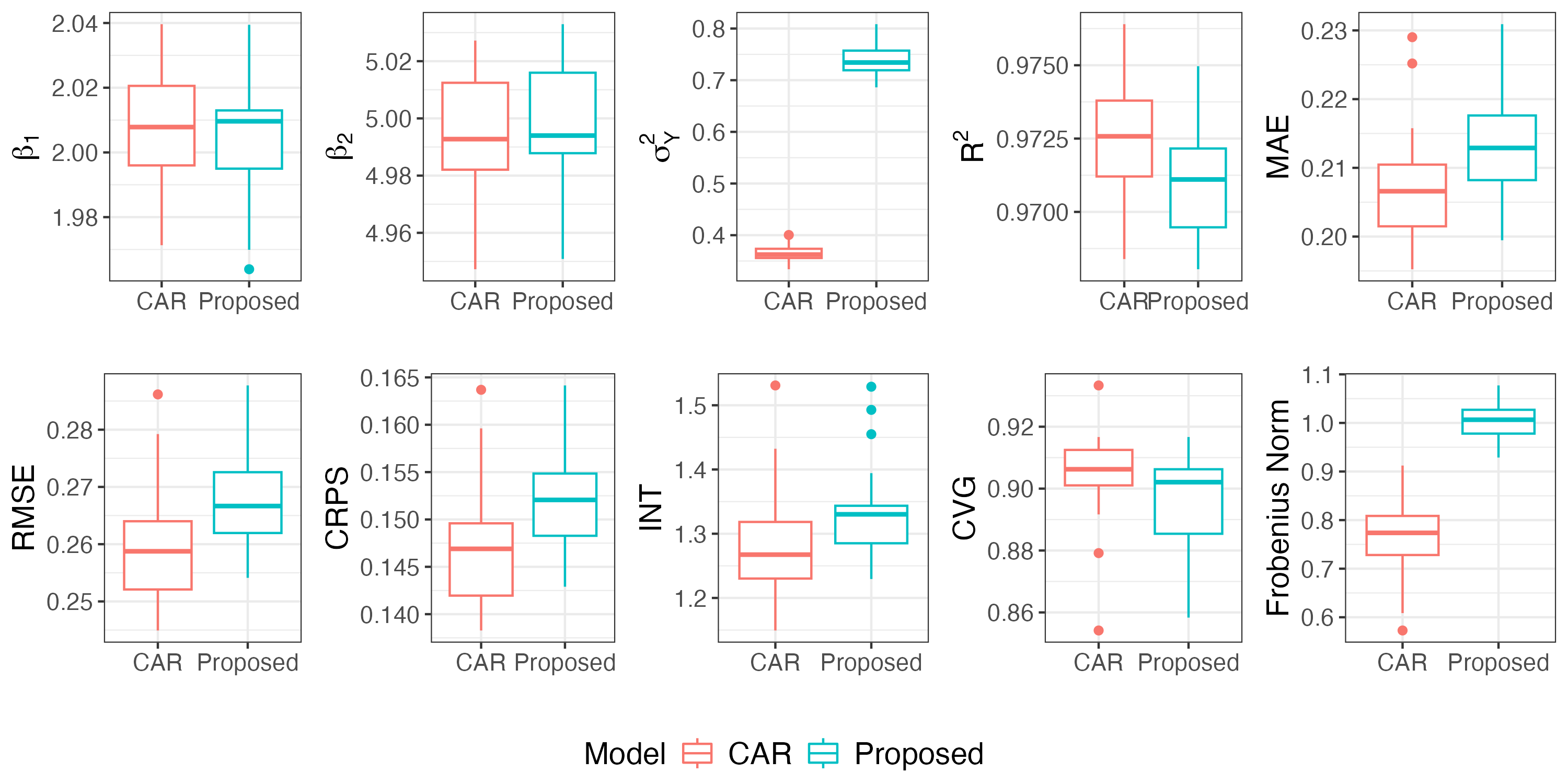}
    \caption{Performance comparison between the TAR\textsubscript{C} model and the CAR model when the data is simulated from the CAR model, i.e., Item (a). For each metric, we plot the Boxplot of 20 replicates.}
    \label{2fig:comparecar}
\end{figure}

\subsection{Comparison with CAR and SAR Model} \label{2subsec:simstudyCarSar}
We now compare the performance of the proposed TAR model with CAR and SAR model over multiple replicates of the data. It is important since the proposed model becomes CAR and SAR model with limiting conditions (Corollaries \ref{2cor:1} and \ref{2cor:3}). To make this comparison, we first simulate 20 replicates of the data using Items (a) and (b) (mentioned in Section \ref{2subsec:illusCarSar}) separately to compare the TAR\textsubscript{C} model with the CAR model in \eqref{2eq:carJoint2}. Then we simulate 20 replicates of the data using Items (c) and (d) (mentioned in Section \ref{2subsec:illusCarSar}) separately to compare the TAR\textsubscript{S} model and the SAR model in \eqref{2eq:sarJoint2}. For predictions at the missing locations, we use the Kriging predictor for both CAR and SAR model. We draw 500 posterior samples directly from the respective posterior distribution (refer to Appendix \ref{2app:c} for the sampler of CAR and SAR).

\begin{figure}[t!]
    \centering
    \includegraphics[scale=0.63]{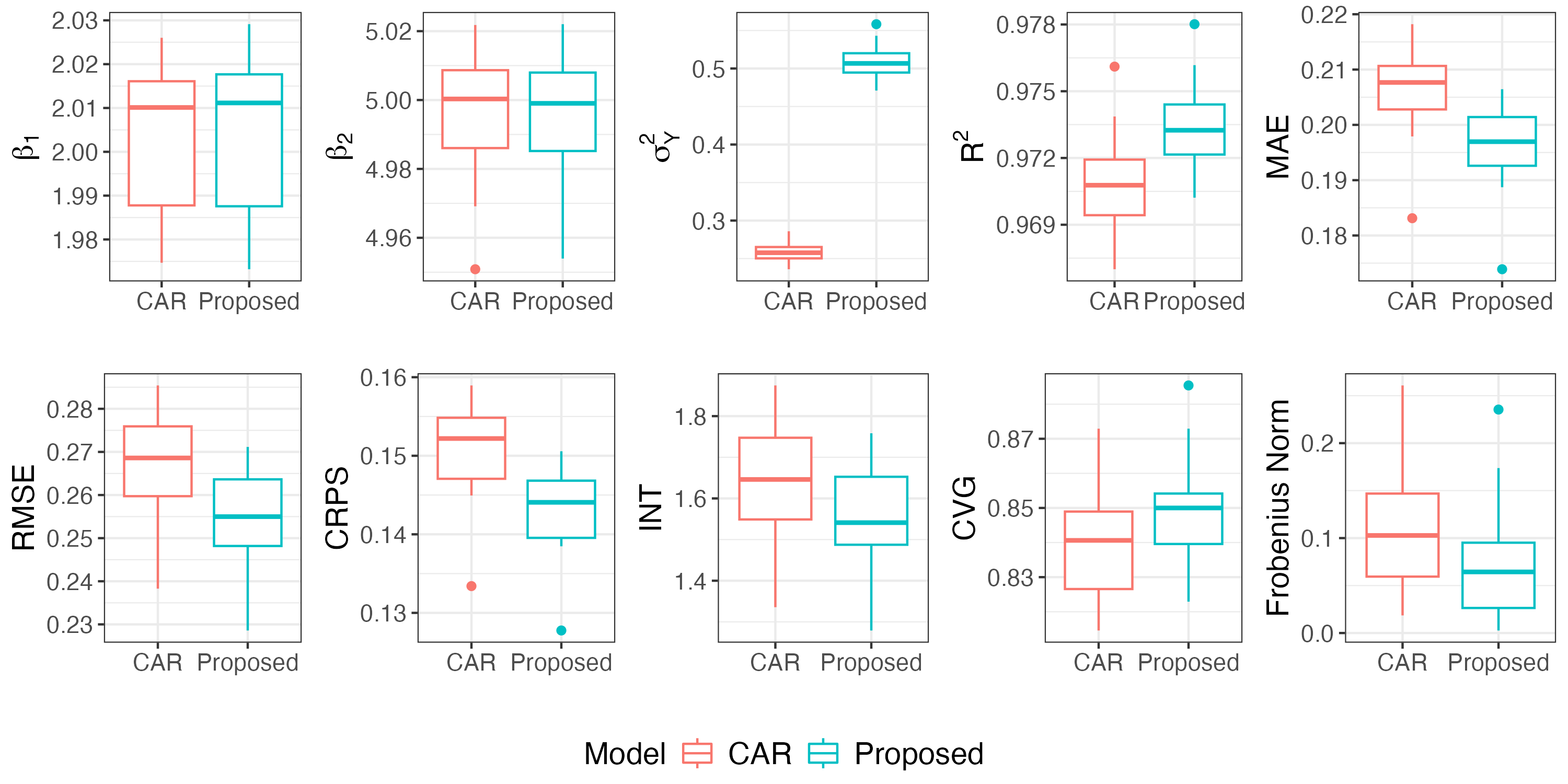}
    \caption{Performance comparison between the TAR\textsubscript{C} model and the CAR model when the data is simulated from the proposed model itself, i.e., Item (b) with $\tau_{Y}^{2}=\sigma_{Y}^{2}$. For each metric, we plot the Boxplot of 20 replicates.}
    \label{2fig:comparetcar}
\end{figure}

In Figure \ref{2fig:comparecar}, we show the performance comparison between TAR\textsubscript{C} and CAR model when both are implemented on the datasets simulated from a CAR model where we consider 20 realizations. The median estimate of $\beta_1$ and $\beta_2$ are almost same for both model, and the median $\sigma_{Y}^{2}$ is closer to its true value 0.5 for the CAR model. We plot the prediction accuracy metrics like $R^{2}$, MAE, RMSE, CRPS, INT and CVG score at $\alpha=0.05$. CRPS measures the closeness of the true data to the predicted distribution, a lower INT score implies that more point predictions are within $(1-\alpha) \times 100\%$ prediction interval with lower and upper predictive quantiles are at levels $\frac{\alpha}{2}$ and $1-\frac{\alpha}{2}$, and CVG calculates the percent of $(1-\alpha) \times 100\%$ prediction intervals containing the true value. In terms of predicting $\bld{y}$ at missing locations, the CAR model does a better job than the TAR\textsubscript{C} model in all the performance metrics. This is expected since the data is generated from the CAR model itself. One thing to notice is that the Boxplots of the prediction accuracy metrics have overlapping values between the two models. This indicates that our model's predictive performance is similar to the CAR model. Practically speaking, the CAR model does not significantly outperform TAR\textsubscript{C} model even when the data is generated from the CAR model, and the proposed model gives a fairly competitive performance. We also calculate the Frobenius norm of the difference between the true and estimated covariance. The Frobenius norm of the difference between $\bld{K}$ and $\bld{L}$ matrices is defined as $\sqrt{\Tr\{(\bld{K}-\bld{L})'(\bld{K}-\bld{L})\}}$ which calculates the sum of squared difference between each element of $\bld{K}$ and $\bld{L}$. Hence, a lower Frobenius norm for the CAR model shows that the estimated covariance is closer to the true covariance than that for the TAR\textsubscript{C} model.

In Figure \ref{2fig:comparetcar}, the performance comparison is drawn between TAR\textsubscript{C} and CAR model when both are implemented on the datasets simulated from a the TAR\textsubscript{C} model itself. Here also we realize 20 replicates of the data generated from the same simulation setting in Item (b) where we set $\delta=1$ per our model (i.e., so the TAR\textsubscript{C} model is correctly specified). In this case, we again see that the median estimates of $\beta_1$ and $\beta_2$ are very similar for both the models. However, the median $\sigma_{Y}^{2}$ is very close to its true value for TAR\textsubscript{C} model. We also notice the although we find overlapping values between the Boxplots, the proposed model outperforms the CAR model in prediction accuracy with respect to all the metrics. The Frobenius norms of the difference between the true and the estimated covariance are lesser for the TAR\textsubscript{C} model that indicates that the proposed model estimates the covariance matrix more accurately. Also, the values itself are smaller than the ones in Figure \ref{2fig:comparecar} which shows that TAR\textsubscript{C} model gives a better estimate of the covariance matrix when the data is generated from the model itself.

\begin{figure}[t!]
    \centering
    \includegraphics[scale=0.63]{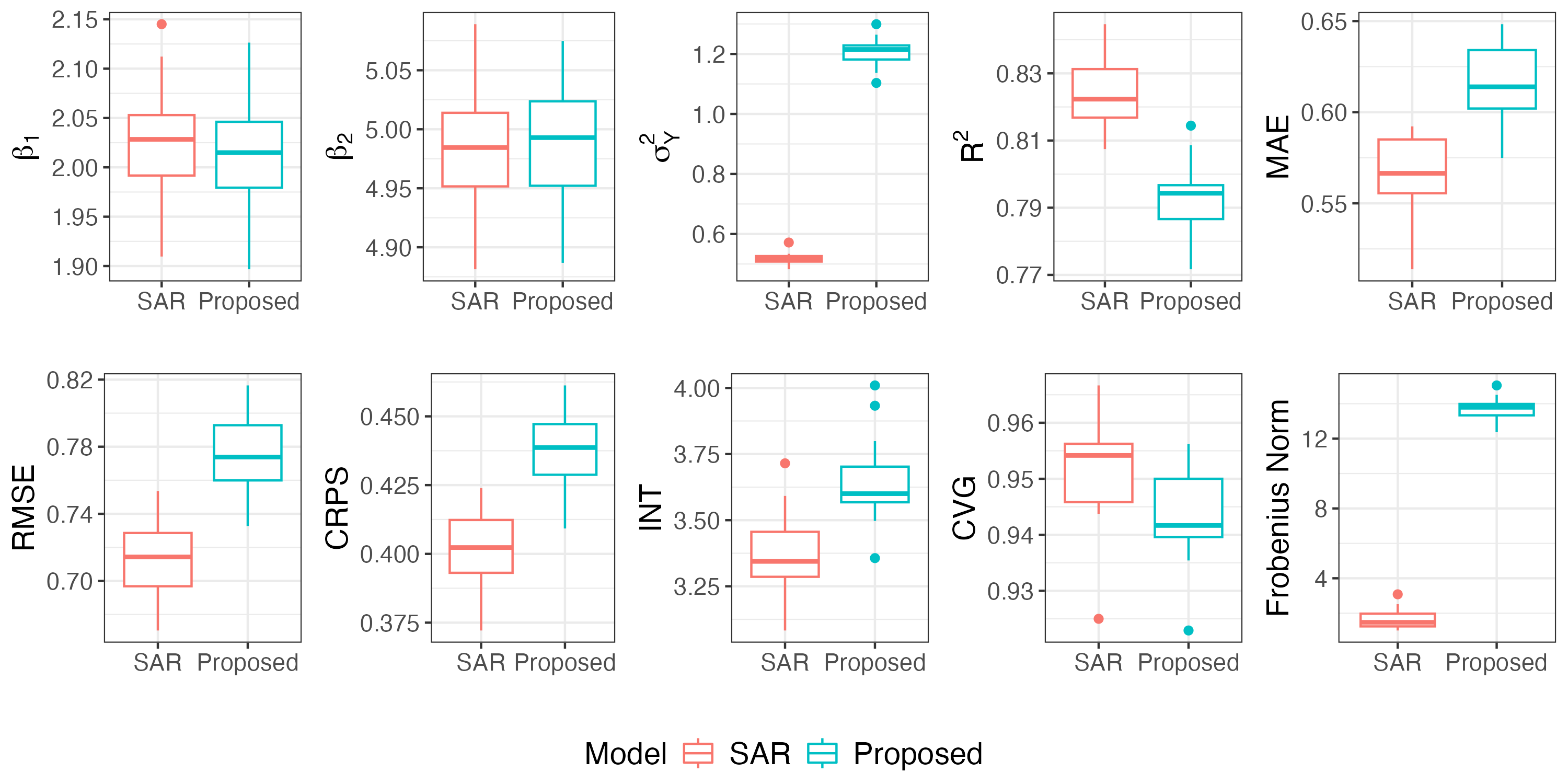}
    \caption{Performance comparison between the TAR\textsubscript{S} model and the SAR model when the data is simulated from the SAR model, i.e., Item (c). For each metric, we plot the Boxplot of 20 replicates.}
    \label{2fig:comparesar}
\end{figure}

In Figure \ref{2fig:comparesar}, we show the performance comparison between TAR\textsubscript{S} and SAR model when implemented on the datasets (20 replicates) simulated from the SAR model itself. Here, the median estimates of both $\beta_{1}$ and $\beta_{2}$ from the proposed model are appeared to be slightly closer to its true value than SAR model, and the SAR is more clearly preferable when it comes to the median $\sigma_{Y}^{2}$. Furthermore, the SAR model shows better accuracy in predicting $\bld{y}$ at the missing locations with respect to all the metrics, with some overlapping regions between the Boxplots. Note that this type of result is expected since the data is generated from the SAR model itself. Based on the Boxplots of the Frobenius norms, SAR model provides a better estimate of the covariance matrix than TAR\textsubscript{S} model.

\begin{figure}[t!]
    \centering
    \includegraphics[scale=0.63]{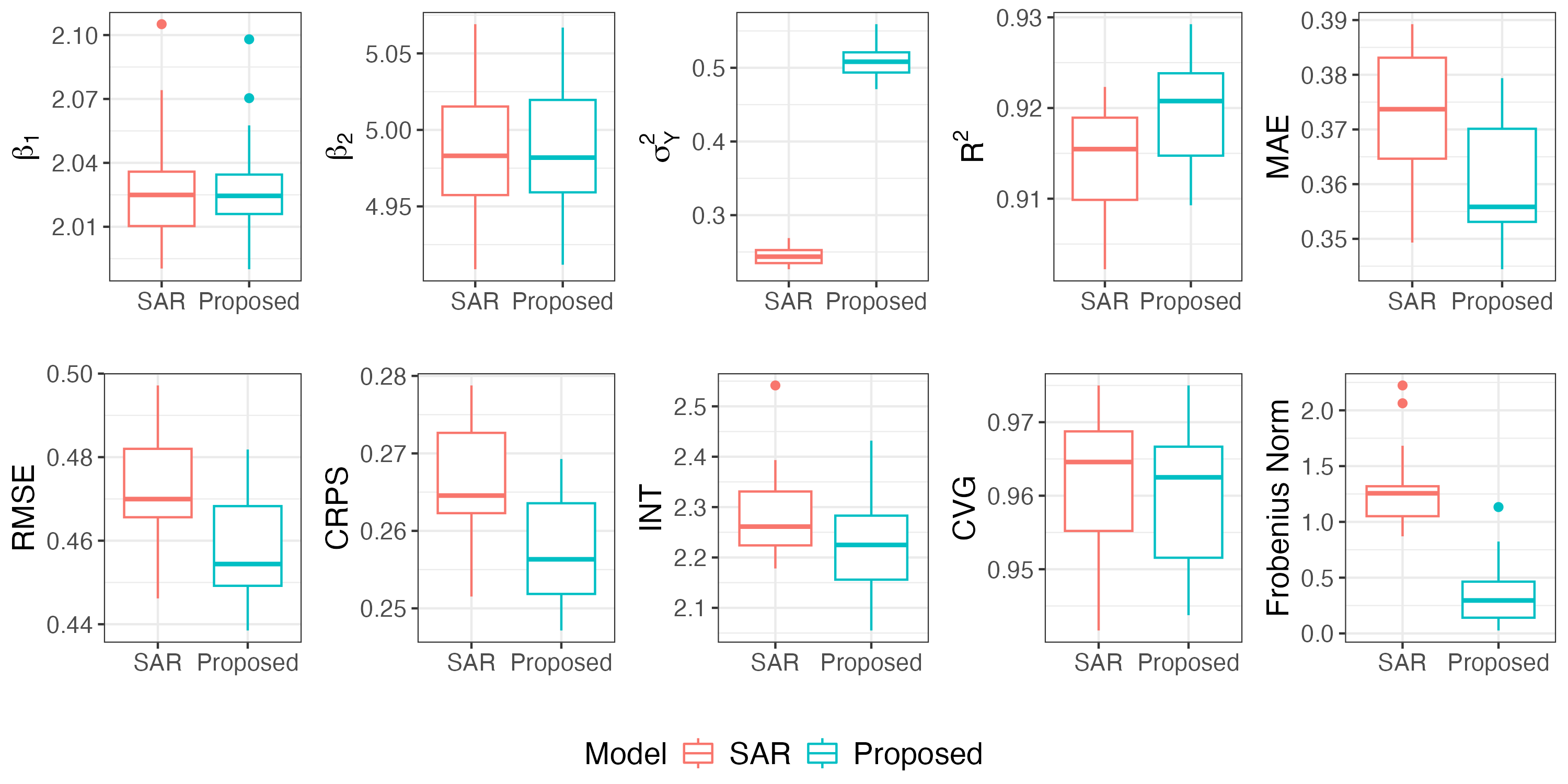}
    \caption{Performance comparison between the TAR\textsubscript{S} model and the SAR model when the data is simulated from the proposed model itself, i.e., Item (d) with $\tau_{Y}^{2}=\sigma_{Y}^{2}$. For each metric, we plot the Boxplot of 20 replicates.}
    \label{2fig:comparetsar}
\end{figure}

Figure \ref{2fig:comparetsar} compares the performance between TAR\textsubscript{S} and SAR model when the data is simulated from the proposed model, i.e., Item (d) where we set $\delta=1$. We see that the median estimates of $\beta_1$ and $\beta_2$ are very similar for both the models, but our model estimates $\sigma_{Y}^{2}$ very precisely. The TAR\textsubscript{S} model clearly outperforms SAR in prediction accuracy in all the metrics, but not significantly. The SAR model does a slightly poor job in this case. The Frobenius norm is lesser for the TAR\textsubscript{S} model which indicates that the proposed model correctly estimates the covariance matrix.

Overall, the proposed TAR model outperforms the traditional models when the data is simulated from the TAR model itself. Also, the traditional models do not significantly outperform the TAR model even when the data is generated from the traditional models. This type of results show that TAR model is very competitive to the traditional models and even performs better when the data follows the model.

\subsection{Computation Time} \label{2subsec:comptime}
The proposed TAR model is computationally efficient because it is implemented using a direct sampler which provides a way to draw posterior samples directly from the posterior distribution. The model also motivates a fast prediction step that leverages Neumann series for matrix inversion (to find covarinace matrix by inverting precision matrix) and samples $\bld{y}_{\mathcal{M}}$ in parallel for a set of missing regions/locations. On the other hand, since the CAR and SAR models are standard, researchers have come up with different versions of them to implement the models efficiently. For the study in Section \ref{2subsec:simstudyCarSar}, we implement the traditional CAR and SAR model in \eqref{2eq:carJoint2} and \eqref{2eq:sarJoint2} using a direct sampler, and we provide predictions using Kriging predictor (see Appendix \ref{2app:c}). But, there are other ways to implement the models more efficiently by using different functions in \texttt{R}. In this section, we compare the computation time of the TAR model with different versions of the CAR and SAR model, and we present the result in Table \ref{2table:simcomptime}. We use the author’s laptop computer with the following specification for this study: Apple M3 Pro 12‑core CPU with 4.06 GHz clock speed and 18 GB RAM.

\begin{table}[t!]
    \renewcommand{\arraystretch}{1.1}
    \centering
    \begin{tabular}{@{\hspace{0.3in}}c@{\hspace{0.5in}}c@{\hspace{0.8in}}c@{\hspace{0.3in}}}
        \hline
        \hline
        \multicolumn{2}{p{2.5in}}{\centering \textbf{Model}} & \textbf{CPU Time (sec.)}\\
        \hline
        \multirow{4}{*}{\centering TAR} & TAR\textsubscript{C} & 5.409\\
        & TAR\textsubscript{S} & 5.260\\
        & TAR\textsubscript{C} with $\delta=1$ & 0.354\\
        & TAR\textsubscript{S} with $\delta=1$ & 0.349\\
        \hline
        \multirow{3}{*}{CAR} & Traditional & 167.027\\
        & \texttt{S.CARleroux} & 12.746\\
        & \texttt{S.CARleroux} with $\rho=1$ & 8.255\\
        \hline
        SAR & Traditional & 221.018\\
        \hline
        \hline
    \end{tabular}
    \caption{Model runtime and prediction time of different versions of the CAR and SAR model in comparison to the proposed TAR model.}
    \label{2table:simcomptime}
\end{table}

In Table \ref{2table:simcomptime}, we show the computation time of the traditional CAR model in \eqref{2eq:carJoint2} with $\mathcal{DU}(-0.99,-0.97,\ldots,0.97,0.99)$ prior on $\rho$, the CAR model by \citet{leroux2000estimation} and the intrinsic CAR (ICAR) model, and compare them with the TAR\textsubscript{C} model with the $\mathcal{DU}(0.1,0.2,\ldots,10)$ prior on $\delta$ and with $\delta=1$. We keep the length of discrete support for $\delta$ (in TAR\textsubscript{C}) and $\rho$ (in CAR) same for a fair comparison. We draw 500 posterior samples for TAR\textsubscript{C} and traditional CAR model. The CAR model by \citet{leroux2000estimation} is implemented using \texttt{S.CARleroux} function from \texttt{CARBayes} package in \texttt{R} with $G=20,000$ and a burn-in of 4,000. We use \texttt{S.CARleroux} to implement the ICAR model as well where we set $\rho=1$ in the function. \texttt{S.CARleroux} takes around 13 and 8 seconds to run the CAR and ICAR model respectively along with the prediction. The traditional CAR model takes only 27 milliseconds to run the model, but it takes around 3 minutes to get the predictions. The TAR\textsubscript{C} model is fastest among all with a model runtime of 21 milliseconds and a prediction time of 5.388 seconds (total 5.409 seconds). TAR model draws samples directly from the posterior distribution (thus, even 200 samples would be sufficient for parameter estimation and predictions, and that would take even lesser CPU time), while methods like \citet{leroux2000estimation} uses a Gibbs sampler and it requires to generate at least 10-20 thousand samples to make sure that the samples come from a stationary distribution. Also, TAR\textsubscript{C} model uses Neumann series for matrix inversion, where it computes $(\bld{I}-\bld{D}_w^{-1}\bld{W})$ and its higher orders only \textit{once}. Whereas, for a traditional CAR model $(\bld{D}_w-\rho\bld{W})^{-1}$ must be computed for each $\rho$ using standard technique. This leads to a considerably fast prediction by the TAR\textsubscript{C}, even when it uses a Kriging predictor.

We also compare the computation time of TAR\textsubscript{S} with the traditional SAR model \eqref{2eq:sarJoint2}. We draw 500 posterior samples. The traditional SAR model takes only 18 milliseconds to run, and the prediction step takes around 4 minutes. The TAR\textsubscript{S} model is considerably faster compared to the traditional SAR with a model runtime of only 11 milliseconds and prediction time of 5.249 seconds (total 5.260 seconds). Again, this is because TAR\textsubscript{S} model computes the covariance matrix by leveraging Neumann series where it calculates $(\bld{I}-\bld{A})'(\bld{I}-\bld{A})$ and its higher orders only \textit{once}. Whereas, the SAR model has to compute $[(\bld{I}-\rho\bld{A})'(\bld{I}-\rho\bld{A})]^{-1}$ for each $\rho$ using standard inversion technique. We also implement a SAR model through Metropolis-Hastings algorithm using \texttt{hom\_sar} function from \texttt{BSPADATA} package in \texttt{R} with $G=5,000$ and a burn-in of 1,000. It takes about 1 hour just to run the model, and it does not have a prediction step associated with the function. We reiterate that the TAR model is computationally efficient because of the fast direct posterior sampling strategy and the fast prediction step that involves matrix inverse using Neumann series and sampling $i$-th element of $\bld{y}_{\mathcal{M}}$ in parallel.

\section{Application: Average Property Price across Greater Glasgow in 2008} \label{2sec:application}
The Scottish Statistics database maintains over 250 datasets that are publicly available in their website. One of the datasets that interests statisticians is the property price. Analyzing average property price has been one of the prime interests for decades due to the volatility of the real estate market. The price of a property has a strong dependency on geographical location/area. But along with location, it also depends on several factors including crime rate, property size, property type, number of grocery store, pharmacy nearby, availability of schools nearby and many more. The proposed TAR model is developed using a well defined Markov random field and models the spatial property of the data, which makes the model perfect to analyze this type of dataset which depends on both locations and other factors. In this section, we present a detailed analysis of the dataset consists of average property prices in 2008 across Greater Glasgow, Scotland. Our aim is to learn the spatial pattern in the data, along with quantifying the effect of the respective ecological factors that affect the property prices. The data is originally available at the Scottish Statistics database (\url{https://statistics.gov.scot/home}), but it is made available in the \texttt{CARBayesdata} package in \texttt{R} as ``\texttt{pricedata}''. This dataset is used in the \texttt{CARBayes} package \citep{lee2016carbayes} in \texttt{R}. We compare the result from the proposed TAR\textsubscript{C} model with the worked out example that fits a CAR model by \citet{leroux2000estimation}.

\begin{figure}[t!]
    \centering
    \includegraphics[scale=0.45]{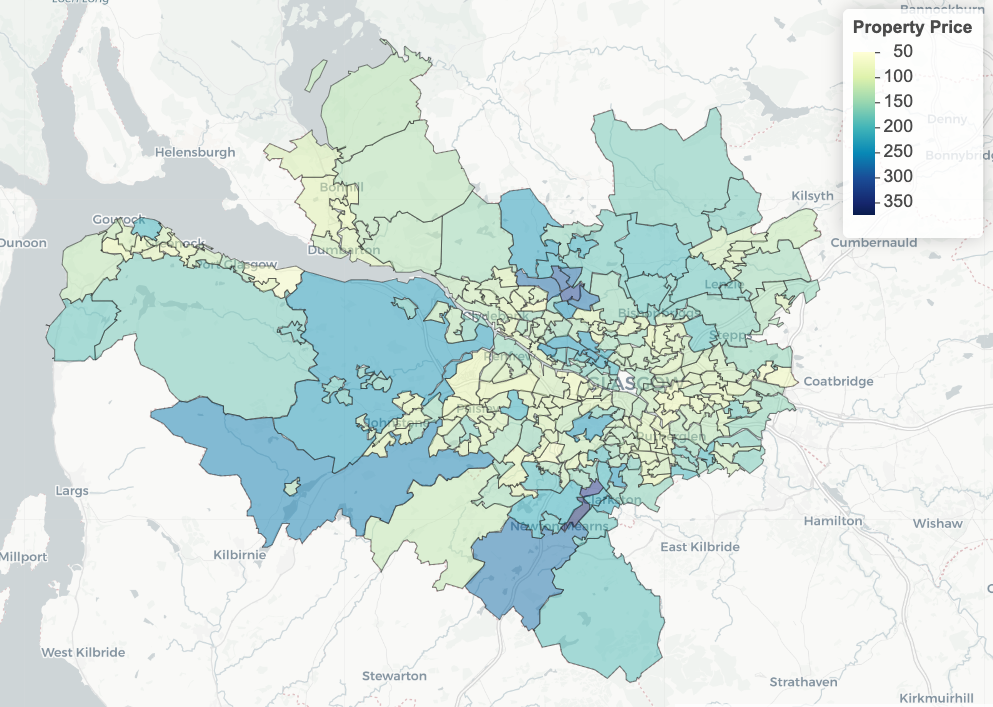}
    \caption{Map view of Greater Glasgow, Scotland with its 270 IZs that show the median property prices in thousand \pounds.}
    \label{2fig:glasgowmap}
\end{figure}

The region of study is the Greater Glasgow and Clyde health board (GGHB) that is divided into 271 small areas with a median population of 4,239. These areas are known as intermediate zones (IZ). The \texttt{pricedata} consists of 270 rows and 7 columns. The rows represent 270 out of 271 IZs in GGHB (one IZ had outlying values which was removed). The 7 columns are -- IZ, price, crime (crime rate), rooms (median number of rooms in a property), sales (number of properties sold in 2008), driveshop (average time taken to drive to the nearest shopping center) and type (most prevalent property type). The response variable here is the ``price'' that represents the median price (in thousand \pounds) of all the properties sold in 2028 in each IZ. The spatial object \texttt{GGHB.IZ} can be loaded form the \texttt{CARBayesdata} package that contains spatial information for the GGHB which helps to show the data in a geographical map and to construct the proximity matrix $\bld{W}$. We show the map of the median property prices at each IZs in the GGHB in Figure \ref{2fig:glasgowmap}.

\begin{table}[t!]
    \renewcommand{\arraystretch}{1.1}
    \centering
    \begin{tabular}{p{0.2\textwidth}cccc}
        \hline
        \hline
        \multirow{2}{*}{\textbf{Coefficients}} & \multirow{2}{*}{\textbf{CARleroux}} & \multicolumn{3}{c}{\textbf{TAR\textsubscript{C}}}\\
        & & $\delta=0.5$ & $\delta=1$ & $\delta=1.5$\\
        \hline
        (Intercept) & 4.134 & 4.222 & 4.181 & 4.153\\
        & (3.864, 4.406) & (3.936, 4.507) & (3.88, 4.482) & (3.865, 4.44)\\
        crime ($\times 10^{-4}$) & -1.4 & -1.5 & -1.4 & -1.3\\
        & (-2.4, -0.5) & (-2.5, -0.6) & (-2.4, -0.4) & (-2.1, -0.3)\\
        rooms & 0.234 & 0.238 & 0.241 & 0.242\\
        & (0.182, 0.284) & (0.185, 0.291) & (0.183, 0.298) & (0.189, 0.295)\\
        sales ($\times 10^{-3}$) & 2.31 & 2.32 & 2.33 & 2.32\\
        & (1.69, 2.94) & (1.63, 3) & (1.68, 2.98) & (1.65, 3)\\
        driveshop ($\times 10^{-2}$) & 0.36 & -3.23 & -2.02 & -1\\
        & (-3.09, 3.82) & (-6.66, 0.2) & (-5.53, 1.49) & (-4.54, 2.54)\\
        type-flat & -0.295 & -0.331 & -0.326 & -0.325\\
        & (-0.407, -0.186) & (-0.449, -0.213) & (-0.442, -0.211) & (-0.437, -0.213)\\
        type-semi & -0.171 & -0.22 & -0.213 & -0.210\\
        & (-0.268, -0.072) & (-0.33, -0.11) & (-0.319, -0.107) & (-0.317, -0.103)\\
        type-terrace & -0.324 & -0.338 & -0.348 & -0.346\\
        & (-0.447, -0.206) & (-0.481, -0.196) & (-0.486, -0.21) & (-0.483, -0.209)\\
        RMSE & 0.233 & 0.226 & 0.227 & 0.229\\
        MAE & 0.185 & 0.181 & 0.181 & 0.182\\
        \hline
        \hline
    \end{tabular}
    \caption{Coefficient estimates (with 95\% credible intervals) for the property prices data as well as RMSE and MAE of the predicted response for TAR\textsubscript{C} and CARleroux model.}
    \label{2table:propertyprice}
\end{table}

The average property price varies between \pounds 50,000 and \pounds 372,750 across the IZs which shows a large variability in the response variable. Also, \citet{lee2016carbayes} has shown that an initial linear regression model (that includes all the covariates) indicates the residual to be non-normal and right skewed. Hence, we transform the the data to the natural log scale and use the transformed response variable in the model. Furthermore, the  property type feature is categorical that records four types of properties -- flat, terrace, semi and detached. We substitute this variable with three dummy variables ``flat'', ``semi'' and ``terrace'' with values 0 or 1 depending on which type of property is prevalent at each IZ (1 means the property type is prevalent and 0 means otherwise). All three dummy variables with 0 indicates that the prevalent property type is detached. We make use of the \texttt{GGHB.IZ} object to find the proximity matrix $\bld{W}$. We implement the TAR\textsubscript{C} model on the natural log of property prices with 7 features and an intercept, and we present the result in Table \ref{2table:propertyprice}. Notice that there is no missing data, hence the model is fitted on all 270 data points. In Table \ref{2table:propertyprice}, the estimates of ``crime'', ``sales'' and ``driveshop'' should be read as the value times $10^{-4}$, $10^{-3}$ and $10^{-2}$ respectively.

\begin{figure}[t!]
    \centering
    \includegraphics[scale=0.45]{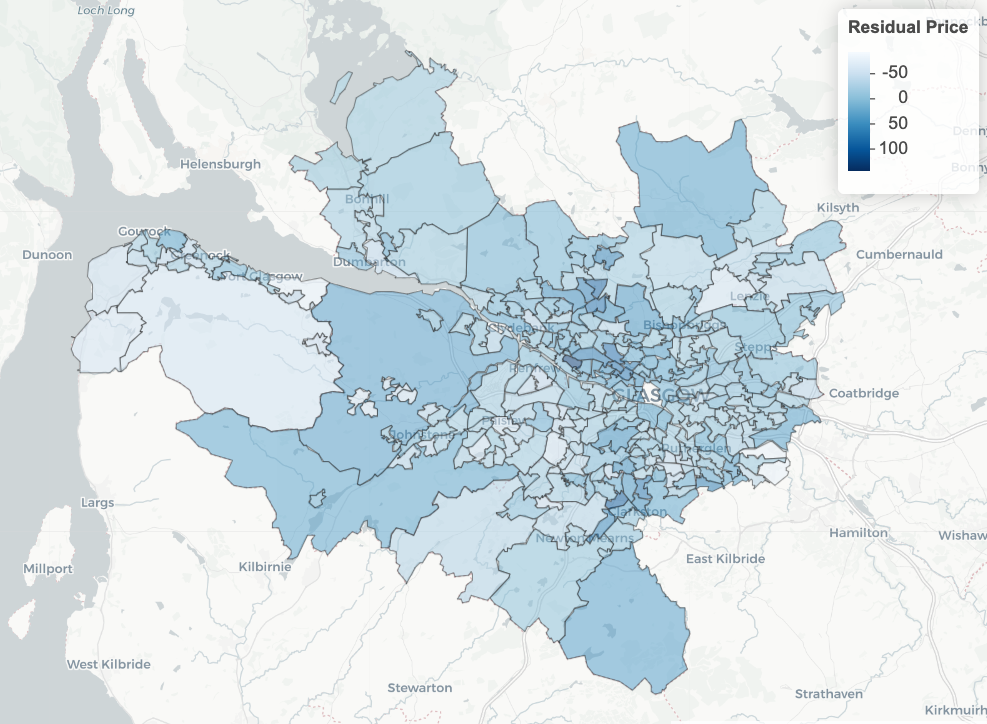}
    \caption{Map view of the residual property prices in thousand {\pounds} when we implement the TAR\textsubscript{C} model with $\delta=1$. A negative price indicates overestimation.}
    \label{2fig:residualprice}
\end{figure}

We implement TAR\textsubscript{C} for $\delta=0.5$, $\delta=1$ and $\delta=1.5$. We observe several interesting results in Table \ref{2table:propertyprice}. Both TAR\textsubscript{C} and CARleroux model have estimated pretty much similar values of the coefficients. There is a small mismatch in type-flat and type-semi. But, notice that driveshop is negative for TAR\textsubscript{C} model while it is positive for CARleroux model. The feature driveshop represents the average time taken to drive to the nearest shopping center. So, the more time one takes to drive, the farther away the property is from the nearest shopping center. Therefore, property prices and driveshop has an inversely proportional relation. TAR\textsubscript{C} model does a fantastic job here in estimating the coefficient to be negative (for all three values of $\delta$) which follows the intuition, whereas the CARleroux model point estimate has a difficult to interpret positive value. Although the 95\% credible interval (from CARleroux) suggests that this coefficient may be zero. Furthermore, both RMSE and MAE between the true and predicted response are lower for TAR\textsubscript{C} model. Considering that TAR\textsubscript{C} and CAR have the same number of parameters (a traditional quantity to adjust for goodness of fit in information criteria), the results suggest that TAR\textsubscript{C} is preferred. The analysis is particularly important for evaluating TAR model to a traditional model. The TAR model has done an excellent job in estimating driveshop that reflects its correct relationship with property prices. It also provides more accurate predictions than CAR with respect to RMSE and MAE. In Figure \ref{2fig:residualprice}, we show the residual property prices at each IZs in the GGHB when implementing TAR\textsubscript{C} with $\delta=1$. A negative residual indicates overestimation and a positive residual indicates underestimation. Most predictions appear to be closer to the observed values which further indicates that the model produced a reasonable fit to the data. 

\section{Discussion} \label{2sec:discussion}
In the spatial statistics literature, CAR and SAR model dominate other methods when it comes to analyzing areal data. While CAR and SAR model define a well established Markov random field, a range parameter $\rho$ is introduced in both of them to make the covariance matrix positive definite. This $\rho$ parameter is defined in a complicated space and can be difficult to learn \citep{wall2004close}. In this article, we propose a novel truncated autoregressive (TAR) model where we enforce the proximity information through truncation, and this eliminates the need of including a $\rho$ parameter into the model. TAR is defined using variance parameters which is much easier to learn and interpret, which is the primary contribution of the article. The proposed TAR model defines a Markov random field and is a competing method to CAR and SAR, and provides an avenue for future methodological development. We define the Markov random fields using truncation applied (i) on the full-conditional distribution at each region (we call this model conditional TAR or TAR\textsubscript{C} model), and (ii) on the joint distribution of the data process (we call this model simultaneous TAR or TAR\textsubscript{S} model). We develop several theoretical results that connect the TAR model to the traditional CAR and SAR model, and to the NNGP model when the precision-nugget $\tau_{Y}^{2}$ is infinite. TAR is implemented using a sampler that draws the posterior samples directly from the posterior distribution which is defined with respect to the precision matrix. Thus, the model possesses a huge computational advantage. Another contribution is that TAR motivates a very fast posterior predictive step that uses the Neumann series to aid with matrix computations.

We demonstrate the TAR model's inferential and predictive performance through simulation studies. We have found that the proposed TAR\textsubscript{C} and TAR\textsubscript{S} model correctly estimate $\bld{\beta}$ and $\{\bld{\beta},\sigma_{Y}^{2}\}$ when the data is generated from a traditional CAR and SAR model and from the proposed model itself, respectively. We also analyze the scatter plot between true and predicted data at missing locations and they appear to be on the 45{\degree} reference line. It implies that the model correctly predicts the data at the missing locations. As expected, the predictions are more accurate when the data is generated from the TAR model. We also make a performance comparison between TAR\textsubscript{C} and CAR model, and between TAR\textsubscript{S} and SAR model when the data is generated from each of these models. We have found that CAR outperforms TAR\textsubscript{C} when the data is generated from CAR model, and it is the other way around when the data is generated from TAR\textsubscript{C} model; however, practically speaking there is little difference in their predictions. We calculate the Frobenius norm of the difference between the true and estimated covariance. It appears to be less for TAR\textsubscript{C} in the latter case. We also see that even when the data is generated from CAR model, TAR\textsubscript{C} model performs almost similarly to the CAR model. Similar results have been found in the case of SAR vs. TAR\textsubscript{S}. In terms of computation time, the proposed TAR model appears to be the fastest among traditional CAR, SAR, and CARleroux, since it is implemented on a direct sampler, and has a fast prediction step that leverages the Neumann series as well as sampling $i$-th element of $\bld{y}_{\mathcal{M}}$ in parallel for a set of missing regions.

In our application, we evaluate the inferential and predictive capability of the proposed TAR model. We have analyzed a property prices dataset that is dependent on both geographical location and ecological factors and compared its performance with the CARleroux model. We have found that the TAR model produces more interpretable coefficients based on their relationship to the response. Particularly, the model has estimated driveshop (one of the features) to be negative whereas CARleroux had estimated this to be positive. Driveshop represents the average time taken to drive to the nearest shopping center. Thus, a lower driveshop naturally results in higher property price and the TAR model estimated a negative coefficient while the CARleroux most estimated a less interpretable positive coefficient. Furthermore, the lower RMSE and MAE suggests that the TAR model have produced more accurate predictions of the property price than CARleroux.

We have assigned a uniform prior to the auxiliary variable $u_{i}$. This establishes the connections of the proposed TAR model to the CAR and SAR model. One can certainly choose a different prior other than uniform for $u_{i}$, and it would lead to a different model. In that case, the proposed model may not have a connection to CAR and SAR. Thus, choosing different prior for $u_{i}$ opens a door to potential future research. Similarly, one might choose to center the constraint around other predictors. In the Appendix, we demonstrate one alternative predictor to define the constraint that leads to an interesting connection with the NNGP.

\newpage
\appendix
\appendixpage
\section{Technical Results of TAR}
In this section, we present the proof of necessary technical results of the TAR model referenced in the main text.

\subsection{Proof of Propositions \ref{2prop:1} and \ref{2prop:2}} \label{2app:a1}
We start with Equation \eqref{2eq:trunc1}, the definition of truncation on each $Y_{i}\vert\bld{y}(N_{i}),\forall i=1,\ldots,n$, and we integrate out $u_{i}$ in the following way.
{\small
\begin{align}
    &f\Bigl(Y_{i}\vert\bld{y}(N_{i}),\bld{\beta},\sigma_{Yi}^{2},\tau_{Yi}^{2}\Bigr) \nonumber \\
    &= \int_{0}^{1} f\Bigl(Y_{i},u_{i}\vert\bld{y}(N_{i}),\bld{\beta},\sigma_{Yi}^{2},\tau_{Yi}^{2}\Bigr) du_{i} \nonumber \\
    &= \int_{0}^{1} f\Bigl(Y_{i}\vert\bld{y}(N_{i}),\bld{\beta},\sigma_{Yi}^{2},\tau_{Yi}^{2},u_{i}\Bigr)\pi(u_{i}) du_{i} \nonumber \\
    &\propto \int_{0}^{1} \exp\left(-\frac{\widetilde{Y}_{i}^{2}}{2\tau_{Yi}^{2}}\right) \mathbbm{1}\left(- \sqrt{-2\sigma_{Yi}^{2}\log u_{i}} < \widetilde{Y}_{i} - \sum_{j \in N_{i}} a_{ij}\widetilde{Y}_{j} < \sqrt{-2\sigma_{Yi}^{2}\log u_{i}}\right)\mathbbm{1}\left(0 < u_{i} < 1\right) du_{i} \nonumber \\
    &= \int_{0}^{1} \exp\left(-\frac{\widetilde{Y}_{i}^{2}}{2\tau_{Yi}^{2}}\right) \mathbbm{1}\left(0 < \frac{\left\{\widetilde{Y}_{i} - \sum_{j \in N_{i}} a_{ij}\widetilde{Y}_{j}\right\}^{2}}{2\sigma_{Yi}^{2}} < -\log u_{i}\right) \mathbbm{1}\left(0 < u_{i} < 1\right) du_{i} \nonumber \\
    &= \exp\left(-\frac{\widetilde{Y}_{i}^{2}}{2\tau_{Yi}^{2}}\right) \int_{0}^{1} \mathbbm{1}\left(0 < u_{i} < \exp \left(-\frac{\left\{\widetilde{Y}_{i} - \sum_{j \in N_{i}} a_{ij}\widetilde{Y}_{j}\right\}^{2}}{2\sigma_{Yi}^{2}}\right)\right) du_{i}, \nonumber \\
    &\implies f\Bigl(Y_{i}\vert\bld{y}(N_{i}),\bld{\beta},\sigma_{Yi}^{2},\tau_{Yi}^{2}\Bigr) \propto \exp\left(-\frac{\widetilde{Y}_{i}^{2}}{2\tau_{Yi}^{2}} - \frac{\left\{\widetilde{Y}_{i} - \sum_{j \in N_{i}} a_{ij}\widetilde{Y}_{j}\right\}^{2}}{2\sigma_{Yi}^{2}}\right), \label{AA1}
\end{align}
}%
where note $\exp \left(-\frac{\left\{\widetilde{Y}_{i} - \sum_{j \in N_{i}} a_{ij}\widetilde{Y}_{j}\right\}^{2}}{2\sigma_{Yi}^{2}}\right) < 1$ since $\frac{-\left\{\widetilde{Y}_{i} - \sum_{j \in N_{i}} a_{ij}\widetilde{Y}_{j}\right\}^{2}}{2\sigma_{Yi}^{2}} < 0$, and we recall $\widetilde{Y}_{i}=Y_{i}-\bld{x}_{i}'\bld{\beta}$. This concludes the proof of Proposition \ref{2prop:1}.

Thus, we have $f\Bigl(Y_{i}\vert\bld{y}(N_{i}),\bld{\beta},\sigma_{Yi}^{2},\tau_{Yi}^{2}\Bigr) \propto \exp\left(-\frac{\widetilde{Y}_{i}^{2}}{2\tau_{Yi}^{2}}\right)f_{CAR}\left(\widetilde{Y}_{i}\vert\widetilde{\bld{y}}(N_{i}),\sigma_{Yi}^{2}\right)$. Now, we apply Brook's lemma \citep{brook1964distinction,besag1974spatial} on \eqref{AA1} to get the corresponding joint distribution. Taking $\bld{X\beta}$ as the reference point for $\widetilde{\bld{y}}$ to be 0, or $\bld{y}$ to be $\bld{X\beta}$ in \eqref{AA1} and applying Brook's Lemma, we get
\begin{align}
    \frac{f\left(\bld{y}\vert\bld{\beta},\{\sigma_{Yi}^{2}\},\{\tau_{Yi}^{2}\}\right)}{f\left(\bld{X\beta}\vert\bld{\beta},\{\sigma_{Yi}^{2}\},\{\tau_{Yi}^{2}\}\right)} &= \prod_{i=1}^{n} \frac{f\left(\widetilde{Y}_{i}\vert 0_{1},\ldots,0_{i-1},\widetilde{Y}_{i+1},\ldots,\widetilde{Y}_{n},\bld{\beta},\{\sigma_{Yi}^{2}\},\{\tau_{Yi}^{2}\}\right)}{f\left(0\vert 0_{1},\ldots,0_{i-1},\widetilde{Y}_{i+1},\ldots,\widetilde{Y}_{n},\bld{\beta},\{\sigma_{Yi}^{2}\},\{\tau_{Yi}^{2}\}\right)} \nonumber \\
    &= \exp\left(-\frac{1}{2}\widetilde{\bld{y}}'\bld{D}_{\tau}^{-1}\widetilde{\bld{y}}\right)f_{CAR}\left(\widetilde{\bld{y}}\vert\{\sigma_{Yi}^{2}\}\right). \label{AA2}
\end{align}
The term $f_{CAR}\left(\widetilde{\bld{y}}\vert\{\sigma_{Yi}^{2}\}\right)$ in \eqref{AA2} is the expression of the CAR model via Brook's lemma leading to
\begin{equation} \label{AA3}
    \bld{y}\vert\bld{\beta},\sigma_{Y}^{2},\tau_{Y}^{2} \sim \mathcal{N}\left(\bld{X\beta},\left[\frac{1}{\tau_{Y}^{2}}\bld{D}_w + \frac{1}{\sigma_{Y}^{2}}(\bld{D}_w-\bld{W})\right]^{-1}\right),
\end{equation}
where we substitute $\bld{D}_{\tau}=\tau_{Y}^{2}\bld{D}_w^{-1}$ and $\bld{D}_{\sigma}=\sigma_{Y}^{2}\bld{D}_w^{-1}$ ($\bld{D}_{\sigma}$ is a diagonal matrix with $(D_{\sigma})_{ii}=\sigma_{Yi}^{2}$) to get the result in \eqref{AA3}. This concludes the proof of Proposition \ref{2prop:2}.

\subsection{Proof of Proposition \ref{2prop:4}} \label{2app:a2}
We start with Equation \eqref{2eq:trunc2}, the definition of truncation on the joint distribution of $Y_{i},\forall i=1,\ldots,n$, and we integrate out the vector $\bld{u} \in \mathcal{D}=(0,1)^{n}$ in the following way.
\begin{align}
    &f\Bigl(\bld{y}\vert \bld{\beta},\sigma_{Y}^{2},\tau_{Y}^{2}\Bigr) \nonumber \\
    &= \int_{\mathcal{D}} f\Bigl(\bld{y},\bld{u}\vert \bld{\beta},\sigma_{Y}^{2},\tau_{Y}^{2}\Bigr) d\bld{u} \nonumber \\
    &= \int_{\mathcal{D}} f\Bigl(\bld{y}\vert \bld{\beta},\sigma_{Y}^{2},\tau_{Y}^{2},\bld{u}\Bigr)\pi(\bld{u}) d\bld{u} \nonumber \\
    &\propto \exp\left(-\frac{1}{2\tau_{Y}^{2}}\widetilde{\bld{y}}'\widetilde{\bld{y}}\right) \prod_{i=1}^{n} \int_{0}^{1} \mathbbm{1}\left(- \sqrt{-2\sigma_{Y}^{2}\log u_{i}} < \widetilde{Y}_{i} - \sum_{j \in N_{i}} a_{ij}\widetilde{Y}_{j} < \sqrt{-2\sigma_{Y}^{2}\log u_{i}}\right) \nonumber \\
    &\qquad \mathbbm{1}(0 < u_{i} < 1) du_{i} \nonumber \\
    &= \exp\left(-\frac{1}{2\tau_{Y}^{2}}\widetilde{\bld{y}}'\widetilde{\bld{y}}\right) \prod_{i=1}^{n} \int_{0}^{1} \mathbbm{1}\left(0 < \frac{\left\{\widetilde{Y}_{i} - \sum_{j \in N_{i}} a_{ij}\widetilde{Y}_{j}\right\}^{2}}{2\sigma_{Y}^{2}} < -\log u_{i}\right) \mathbbm{1}(0 < u_{i} < 1) du_{i} \nonumber \\
    &= \exp\left(-\frac{1}{2\tau_{Y}^{2}}\widetilde{\bld{y}}'\widetilde{\bld{y}}\right) \prod_{i=1}^{n} \int_{0}^{1} \mathbbm{1}\left(0 < u_{i} < \exp \left(-\frac{\left\{\widetilde{Y}_{i} - \sum_{j \in N_{i}} a_{ij}\widetilde{Y}_{j}\right\}^{2}}{2\sigma_{Y}^{2}}\right) \right) du_{i} \nonumber \\
    &= \exp\left(-\frac{1}{2\tau_{Y}^{2}}\widetilde{\bld{y}}'\widetilde{\bld{y}}\right) \prod_{i=1}^{n} \exp\left(- \frac{\left\{\widetilde{Y}_{i} - \sum_{j \in N_{i}} a_{ij}\widetilde{Y}_{j}\right\}^{2}}{2\sigma_{Y}^{2}}\right) \nonumber \\
    &= \exp\left(-\frac{1}{2\tau_{Y}^{2}}\widetilde{\bld{y}}'\widetilde{\bld{y}}\right) \exp\left(- \frac{\sum_{i=1}^{n}\left\{\widetilde{Y}_{i} - \sum_{j \in N_{i}} a_{ij}\widetilde{Y}_{j}\right\}^{2}}{2\sigma_{Y}^{2}}\right) \nonumber \\
    &= \exp\left(-\frac{1}{2\tau_{Y}^{2}}\widetilde{\bld{y}}'\widetilde{\bld{y}}\right) \exp\left(- \frac{1}{2\sigma_{Y}^{2}}(\widetilde{\bld{y}}-\bld{A\widetilde{Y}})'(\widetilde{\bld{y}}-\bld{A\widetilde{Y}})\right), \nonumber \\
    &= \exp\left(-\frac{1}{2}(\bld{y}-\bld{X\beta})'\left[\frac{1}{\tau_{Y}^{2}}\bld{I} + \frac{1}{\sigma_{Y}^{2}}(\bld{I}-\bld{A})'(\bld{I}-\bld{A})\right](\bld{y}-\bld{X\beta})\right) \nonumber \\
    &\implies \bld{y}\vert \bld{\beta},\sigma_{Y}^{2},\tau_{Y}^{2} \sim \mathcal{N}\left(\bld{X\beta},\left[\frac{1}{\tau_{Y}^{2}}\bld{I} + \frac{1}{\sigma_{Y}^{2}}(\bld{I}-\bld{A})'(\bld{I}-\bld{A})\right]^{-1}\right) \label{AA4}
\end{align}
This concludes the proof of Proposition \ref{2prop:4}.

\subsection{Posterior Distribution of TAR} \label{2app:a3}
Now, we find the posterior distribution from the hierarchical model in \eqref{2eq:BHM}. First, we derive the conditional distribution of $\bld{\beta}$ given $\sigma_{Y}^{2}$, $\delta$ and $\bld{y}_{O}$.
\begin{align}
    f(\bld{\beta}\vert\sigma_{Y}^{2},\delta,\bld{y}_{O}) &\propto p(\bld{y}_{O}\vert\bld{\beta},\sigma_{Y}^{2},\delta)\pi(\bld{\beta}) \nonumber \\
    &\propto \exp\left(-\frac{1}{2\sigma_{Y}^{2}}(\bld{y}_{O}-\bld{X}_{O}\bld{\beta})'(\bld{O}\bld{\Sigma}(\delta)^{-1}\bld{O}')(\bld{y}_{O}-\bld{X}_{O}\bld{\beta})\right) \times 1 \nonumber \\
    &\propto \exp\left(-\frac{1}{2\sigma_{Y}^{2}}\{\bld{\beta}'\bld{X}_{O}'(\bld{O}\bld{\Sigma}(\delta)^{-1}\bld{O}')\bld{X}_{O}\bld{\beta} - 2\bld{\beta}'\bld{X}_{O}'(\bld{O}\bld{\Sigma}(\delta)^{-1}\bld{O}')\bld{y}_{O}\}\right), \nonumber \\
    \implies \bld{\beta}\vert\sigma_{Y}^{2},\delta,\bld{y}_{O} &\sim \mathcal{N}\left((\bld{X}_{O}'\bld{\Sigma}_{o}(\delta)^{-1}\bld{X}_{O})^{-1}\bld{X}_{O}'\bld{\Sigma}_{o}(\delta)^{-1}\bld{y}_{O},\sigma_{Y}^{2}(\bld{X}_{O}'\bld{\Sigma}_{o}(\delta)^{-1}\bld{X}_{O})^{-1}\right), \label{AA6}
\end{align}
where $\bld{\Sigma}_{o}(\delta)^{-1}=\bld{O}\bld{\Sigma}(\delta)^{-1}\bld{O}'$. The last line in \eqref{AA6} is achieved by completing the squares. We use the distribution of $\bld{\beta}\vert\sigma_{Y}^{2},\delta,\bld{y}_{O}$ to derive the joint distribution of $\sigma_{Y}^{2}$, $\delta$ and $\bld{y}_{O}$. We get
\begin{equation} \label{AA7}
    f(\sigma_{Y}^{2},\delta,\bld{y}_{O}) = \frac{f(\bld{\beta},\sigma_{Y}^{2},\delta,\bld{y}_{O})}{f(\bld{\beta}\vert\sigma_{Y}^{2},\delta,\bld{y}_{O})} = \frac{f(\bld{y}_{O}\vert\bld{\beta},\sigma_{Y}^{2},\delta)\pi(\bld{\beta})\pi(\sigma_{Y}^{2})\pi(\delta)}{f(\bld{\beta}\vert\sigma_{Y}^{2},\delta,\bld{y}_{O})},
\end{equation}
where $\pi(\bld{\beta})$, $\pi(\sigma_{Y}^{2})$ and $\pi(\delta)$ are the prior distributions for $\bld{\beta}$, $\sigma_{Y}^{2}$ and $\delta$ respectively. We denote $\bld{\Sigma}_{\beta}^{-1}=(1/\sigma_{Y}^{2})\bld{X}_{O}'\bld{\Sigma}_{o}(\delta)^{-1}\bld{X}_{O}$ and $\bld{\mu}_{\beta}=(\bld{X}_{O}'\bld{\Sigma}_{o}(\delta)^{-1}\bld{X}_{O})^{-1}\bld{X}_{O}'\bld{\Sigma}_{o}(\delta)^{-1}\bld{y}_{O}$ in \eqref{AA6}. Putting the expressions for likelihood and prior, and substituting \eqref{AA6} into \eqref{AA7}, we get
\begin{align}
    &f(\sigma_{Y}^{2},\delta,\bld{y}_{O}) \nonumber \\
    &\propto \frac{(1/\sigma_{Y}^{2})^{n_{O}/2}\lvert \bld{\Sigma}_{o}(\delta)^{-1}\rvert^{1/2} \exp\left(-\frac{1}{2\sigma_{Y}^{2}}(\bld{y}_{O}-\bld{X}_{O}\bld{\beta})'\bld{\Sigma}_{o}(\delta)^{-1}(\bld{y}_{O}-\bld{X}_{O}\bld{\beta})\right)}{\lvert \bld{\Sigma}_{\beta}^{-1}\rvert^{1/2} \exp\left(-\frac{1}{2}(\bld{\beta}-\bld{\mu}_{\beta})'\bld{\Sigma}_{\beta}^{-1}(\bld{\beta}-\bld{\mu}_{\beta})\right)} \hspace{5pt} \pi(\sigma_{Y}^{2})\pi(\delta) \nonumber \\
    &= \frac{(1/\sigma_{Y}^{2})^{n_{O}/2}\lvert \bld{\Sigma}_{o}(\delta)^{-1}\rvert^{1/2}\exp\left(-\frac{1}{2\sigma_{Y}^{2}}\bld{y}_{O}'\bld{\Sigma}_{o}(\delta)^{-1}\bld{y}_{O}\right)}{(1/\sigma_{Y}^{2})^{p/2}\lvert \bld{X}_{O}'\bld{\Sigma}_{o}(\delta)^{-1}\bld{X}_{O}\rvert^{1/2}\exp\left(-\frac{1}{2}\bld{\mu}_{\beta}'\bld{\Sigma}_{\beta}^{-1}\bld{\mu}_{\beta}\right)} \left(\frac{1}{\sigma_{Y}^{2}}\right)^{a+1} \exp\left(-\frac{b}{\sigma_{Y}^{2}}\right) \pi(\delta) \nonumber \\
    &=\lvert \bld{\Sigma}_{o}(\delta)^{-1}\rvert^{1/2} \lvert \bld{X}_{O}'\bld{\Sigma}_{o}(\delta)^{-1}\bld{X}_{O}\rvert^{-1/2} \pi(\delta) \left(\frac{1}{\sigma_{Y}^{2}}\right)^{a+\frac{n_{O}-p}{2}+1} \nonumber \\
    &\quad \exp\left(-\frac{1}{\sigma_{Y}^{2}}\left[b+\frac{1}{2}\left\{\bld{y}_{O}'\bld{\Sigma}_{o}(\delta)^{-1}\bld{y}_{O} - \bld{y}_{O}'\bld{\Sigma}_{o}(\delta)^{-1}\bld{X}_{O}(\bld{X}_{O}'\bld{\Sigma}_{o}(\delta)^{-1}\bld{X}_{O})^{-1}\bld{X}_{O}'\bld{\Sigma}_{o}(\delta)^{-1}\bld{y}_{O}\right\}\right]\right) \label{AA8}.
\end{align}
Substituting \eqref{AA8} into the following expression and we get
\begin{align}
    &f(\sigma_{Y}^{2}\vert\delta,\bld{y}_{O}) \propto f(\sigma_{Y}^{2},\delta,\bld{y}_{O}) \propto \left(\frac{1}{\sigma_{Y}^{2}}\right)^{a+\frac{n_{O}-p}{2}+1} \nonumber \\
    &\quad \exp\left(-\frac{1}{\sigma_{Y}^{2}}\left[b+\frac{1}{2}\left\{\bld{y}_{O}'\bld{\Sigma}_{o}(\delta)^{-1}\bld{y}_{O} - \bld{y}_{O}'\bld{\Sigma}_{o}(\delta)^{-1}\bld{X}_{O}(\bld{X}_{O}'\bld{\Sigma}_{o}(\delta)^{-1}\bld{X}_{O})^{-1}\bld{X}_{O}'\bld{\Sigma}_{o}(\delta)^{-1}\bld{y}_{O}\right\}\right]\right) \nonumber \\
    &\implies \sigma_{Y}^{2}\vert\delta,\bld{y}_{O} \sim \nonumber \\
    &\mathcal{IG}\left(a+\frac{n_{O}-p}{2},b+\frac{1}{2}\left[\bld{y}_{O}'\bld{\Sigma}_{o}(\delta)^{-1}\bld{y}_{O} - \bld{y}_{O}'\bld{\Sigma}_{o}(\delta)^{-1}\bld{X}_{O}(\bld{X}_{O}'\bld{\Sigma}_{o}(\delta)^{-1}\bld{X}_{O})^{-1}\bld{X}_{O}'\bld{\Sigma}_{o}(\delta)^{-1}\bld{y}_{O}\right]\right). \label{AA9}
\end{align}
Now, we find the joint distribution of $\delta$ and $\bld{y}_{O}$ to find $f(\rho\vert\bld{y}_{O})$. We write
\begin{align}
    &f(\delta,\bld{y}_{O}) = \frac{f(\sigma_{Y}^{2},\delta,\bld{y}_{O})}{f(\sigma_{Y}^{2}\vert \rho,\bld{y}_{n_{1}})} \nonumber \\
    &= \frac{\lvert \bld{\Sigma}_{o}(\delta)^{-1}\rvert^{1/2} \lvert \bld{X}_{O}'\bld{\Sigma}_{o}(\delta)^{-1}\bld{X}_{O}\rvert^{-1/2} \mathbbm{1}(\delta_{1},\ldots,\delta_{k})}{\left[b+\frac{1}{2}\left\{\bld{y}_{O}'\bld{\Sigma}_{o}(\delta)^{-1}\bld{y}_{O} - \bld{y}_{O}'\bld{\Sigma}_{o}(\delta)^{-1}\bld{X}_{O}(\bld{X}_{O}'\bld{\Sigma}_{o}(\delta)^{-1}\bld{X}_{O})^{-1}\bld{X}_{O}'\bld{\Sigma}_{o}(\delta)^{-1}\bld{y}_{O}\right\}\right]^{a + \frac{n_{O}-p}{2}}}, \label{AA10}
\end{align}
where we substitute $f(\sigma_{Y}^{2},\delta,\bld{y}_{O})$ from \eqref{AA8} and $f(\sigma_{Y}^{2}\vert \delta,\bld{y}_{O})$ with its expression to get \eqref{AA10}. $\pi(\delta)$ is the mass function of discrete uniform distribution with support $\{\delta_{1},\ldots,\delta_{k}\}$. Based on the definition of conditional distribution, we get
\begin{equation} \label{AA11}
    f(\delta\vert \bld{y}_{O}) = \frac{f(\delta,\bld{y}_{O})}{\sum_{\delta \in \{\delta_{1},\ldots,\delta_{k}\}}f(\delta,\bld{y}_{O})},
\end{equation}
where $f(\delta,\bld{y}_{O})$ is defined in \eqref{AA10}. We calculate the probability mass for each $\{\delta_{1},\ldots,\delta_{k}\}$ and we draw posterior samples of $\delta$ from $\{\delta_{1},\ldots,\delta_{k}\}$ with the respective probability mass. Equations \eqref{AA6}, \eqref{AA9} and \eqref{AA11} together conclude the result in \eqref{2eq:postdist}.

\section{A Connection between TAR and NNGP} \label{2app:b}
The NNGP \citep{datta2016nngp,datta2022nearest} and the general Vecchia \citep{vecchia1988estimation,katzfuss2021general} perspective are important recent contributions that provide one solution to the natural computational difficulties in standard geostatistical models for point-referenced data. In this section, we review the NNGP model, followed by a detailed analysis to establish relationship between TAR and NNGP.

\subsection{A Review of Nearest-Neighbor Gaussian Process Model} \label{2subsec:reviewNNGP}
Nearest-neighbor Gaussian process (NNGP; \citealp{datta2016nngp}) is a very popular fully Bayesian model. It approximates the process model using a smaller conditioning set generated from a directed acyclic graph (DAG) in the following way.
\begin{align} \label{2eq:nngpApprox}
    p(\bld{w}) &= p(w(\bld{s}_1))p(w(\bld{s}_2)\vert w(\bld{s}_1))\ldots p(w(\bld{s}_n)\vert w(\bld{s}_{n-1}),\ldots,w(\bld{s}_1)) \nonumber \\
    &\approx \prod_{i=1}^n p(w(\bld{s}_i)\vert \bld{w}_{S_i}) = \widetilde{p}(\bld{w}),
\end{align}
where the elements of $\bld{w}=(w(\bld{s}_1),\ldots,w(\bld{s}_n))'$ are a spatial process evaluated at $\bld{s}_{1},\ldots,\bld{s}_{n} \in D \in \mathbb{R}^{d}$, $p(\cdot)$ is used for distributions of point-referenced data, and $S_i$ is the set of directed neighbors of $\bld{s}_i$ generated from a DAG (see \citealp{datta2016nngp} for a detailed description on DAG). NNGP considers the distribution of $w(\bld{s}_i)\vert \bld{w}_{S_i}$ to be
\begin{equation} \label{2eq:nngpCond}
    w(\bld{s}_i)\vert \bld{w}_{S_i} \sim \mathcal{N}\left(\bld{b}_{i,S_i}\bld{w}_{S_i},\sigma^{2}f_{\bld{s}_i}\right),
\end{equation}
where $\bld{b}_{i,S_i}=\bld{\mathcal{C}}(\bld{s}_i,S_i)\bld{\mathcal{C}}(S_i,S_i)^{-1}$, $f_{\bld{s}_i}=\mathcal{C}(\bld{s}_i,\bld{s}_i)-\bld{\mathcal{C}}(\bld{s}_i,S_i)\bld{\mathcal{C}}(S_i,S_i)^{-1}\bld{\mathcal{C}}(S_i,\bld{s}_i)$ and $\sigma^{2}$ is the variance. Here, $\bld{\mathcal{C}}$ is a valid cross-correlation function that specifies the true spatial process (e.g., $\bld{\mathcal{C}}$ follows an exponential covariogram with range $\phi$). If the whole process is written as $\bld{w}=\bld{Bw}+\bld{\eta}$, where $\bld{B}=(b_{ij})$ and $\bld{\eta} \sim \mathcal{N}(\bld{0},\sigma^{2}\bld{F})$ with $\bld{F}=\mathrm{diag}(f_{\bld{s}_1},\ldots,f_{\bld{s}_n})$, then from \eqref{2eq:nngpCond} we get
\begin{equation} \label{2eq:nngpJoint}
    \bld{w} \sim \mathcal{N}\left(\bld{0},\sigma^{2}(\bld{I}-\bld{B})^{-1}\bld{F}\left((\bld{I}-\bld{B})^{-1}\right)'\right),
\end{equation}
and $\bld{\mathcal{C}}$ is now approximated with $\widetilde{\bld{\mathcal{C}}}=(\bld{I}-\bld{B})^{-1}\bld{F}\left((\bld{I}-\bld{B})^{-1}\right)'$. Notice that, the structure of covariance matrix for NNGP and SAR model are very similar (refer to Equations \eqref{2eq:sarJoint1} and \eqref{2eq:nngpJoint} for a comparison) with the difference in the definitions of $\bld{B}$, $\bld{D}$ and $\bld{F}$. A detailed discussion on NNGP can be found in \citet{datta2016nngp} and \citet{datta2022nearest}.

\subsection{TAR and NNGP} \label{2subsec:nngpTrunc}
In this section, we introduce a random process into our notation and let the data $\bld{y}=(Y(\bld{s}_{1}),\ldots,Y(\bld{s}_{n}))'$, and $S_{i} \in D \in \mathbb{R}^{d}$. We write the TAR\textsubscript{S} model in a more generic way, where we let $\sigma_{Y}^{2}$ be vary with location by a factor of $f_{i}$, we substitute $a_{ij}$ with $b_{ij} \in \left[0,\infty\right)$, and $N_{i}$ with a generic neighborhood structure $S_{i}$ in the following way:
\begin{align}
    &f\Bigl(\bld{y}\vert \bld{\beta},\sigma_{Y}^{2},\tau_{Y}^{2},\bld{u}\Bigr) \propto \left(\frac{1}{2\pi \tau_{Y}^{2}}\right)^{n/2} \exp\left(-\frac{1}{2\tau_{Y}^{2}}(\bld{y}-\bld{X\beta})'(\bld{y}-\bld{X\beta})\right) \nonumber \\
    &\qquad \qquad \prod_{i=1}^n \mathbbm{1}\left(- \sqrt{-2\sigma_{Y}^{2}f_{i}\log u(\bld{s}_i)} < \widetilde{Y}(\bld{s}_{i}) - \sum_{j:\bld{s}_{j} \in S_{i}} b_{ij}\widetilde{Y}(\bld{s}_{j}) < \sqrt{-2\sigma_{Y}^{2}f_{i}\log u(\bld{s}_{i})}\right), \label{2eq:trunc3}
\end{align}
where the mean adjusted $Y(\bld{s}_i)$ is $\widetilde{Y}(\bld{s}_i)=Y(\bld{s}_i)-\bld{x}(\bld{s}_i)'\bld{\beta}$, $u(\bld{s}_{i}) \stackrel{iid}{\sim} \mathcal{U}(0,1)$ is an auxiliary variable, and $\sigma_{Y}^{2}$ and $\tau_{Y}^{2}$ are variance parameters. Suppose, we let $b_{ij}$ and $f_{i}$ to be dependent on their respective neighborhoods in the following way. Suppose, $\bld{\mathcal{C}}$ is a true correlation function (e.g., exponential, spherical, Mat{\'e}rn etc. indexed by a decay/range parameter $\phi$). We define $S_i$ to be the set of neighborhood locations of $\bld{s}_i$ from a directed acyclic graph from a lattice $\{\bld{s}_{1},\ldots,\bld{s}_{n}\}$ with $S_{i} \in D \in \mathbb{R}^{d}$ which we interpret as different from $N_{i}$ for regional data (refer to \citealp{datta2016nngp} for directed neighbors). Set $\bld{b}_{i,S_i}'=\bld{\mathcal{C}}(\bld{s}_i,S_i)\bld{\mathcal{C}}(S_i,S_i)^{-1}$, and $f_i=\mathcal{C}(\bld{s}_i,\bld{s}_i)-\bld{b}_{i,S_i}'\bld{\mathcal{C}}(S_i,\bld{s}_i)$. We define $\bld{B}=\left(b_{ij}\right)$ to be a lower triangular matrix with non-zero entries at $(i,j): i=1,\ldots,n$ and $j=\{j:\bld{s}_j \in S_i\}$, and $\bld{F}$ to be a diagonal matrix with $F_{ii}=f_i$.

\setcounter{proposition}{0}
\renewcommand{\theproposition}{\Alph{section}\arabic{proposition}}
\begin{proposition} \label{2prop:6}
    Upon integrating out $\{u(\bld{s}_{i})\}$, the model in \eqref{2eq:trunc3} becomes
    \begin{equation} \label{2eq:condY3}
        \bld{y}\vert \bld{\beta},\sigma_{Y}^{2},\tau_{Y}^{2} \sim \mathcal{N}\left(\bld{X\beta},\left[\frac{1}{\tau_{Y}^{2}}\bld{I} + \frac{1}{\sigma_{Y}^{2}}(\bld{I}-\bld{B})'\bld{F}^{-1}(\bld{I}-\bld{B})\right]^{-1}\right),
    \end{equation}
    where $\bld{B}$ is a lower triangular matrix with $\bld{b}_{i,S_i}'=\bld{\mathcal{C}}(\bld{s}_i,S_i)\bld{\mathcal{C}}(S_i,S_i)^{-1}$, $\bld{F}$ is a diagonal matrix with $F_{ii}=\mathcal{C}(\bld{s}_i,\bld{s}_i)-\bld{b}_{i,S_i}'\bld{\mathcal{C}}(S_i,\bld{s}_i)$, and $\bld{\mathcal{C}}=(\mathcal{C}_{ij}(\phi))$ is the true correlation function.
\end{proposition}
\textit{Proof:} We start with Equation \eqref{2eq:trunc3} and we integrate out the vector $\bld{u} \in \mathcal{D}=(0,1)^{n}$ in the following way.
{\small
\begin{align}
    &f\Bigl(\bld{y}\vert \bld{\beta},\sigma_{Y}^{2},\tau_{Y}^{2}\Bigr) \nonumber \\
    &= \int_{\mathcal{D}} f\Bigl(\bld{y},\bld{u}\vert \bld{\beta},\sigma_{Y}^{2},\tau_{Y}^{2}\Bigr) d\bld{u} \nonumber \\
    &= \int_{\mathcal{D}} f\Bigl(\bld{y}\vert \bld{\beta},\sigma_{Y}^{2},\tau_{Y}^{2},\bld{u}\Bigr)\pi(\bld{u}) d\bld{u} \nonumber \\
    &\propto \exp\left(-\frac{1}{2\tau_{Y}^{2}}\widetilde{\bld{y}}'\widetilde{\bld{y}}\right) \prod_{i=1}^{n} \int_{0}^{1} \mathbbm{1}\left(- \sqrt{-2\sigma_{Y}^{2}f_{i}\log u(\bld{s}_i)} < \widetilde{Y}(\bld{s}_i) - \sum_{j:\bld{s}_j \in S_i} b_{ij}\widetilde{Y}(\bld{s}_j) < \sqrt{-2\sigma_{Y}^{2}f_{i}\log u(\bld{s}_i)}\right) \nonumber \\
    &\qquad \mathbbm{1}(0 < u(\bld{s}_i) < 1) du(\bld{s}_i) \nonumber \\
    &= \exp\left(-\frac{1}{2\tau_{Y}^{2}}\widetilde{\bld{y}}'\widetilde{\bld{y}}\right) \prod_{i=1}^{n} \int_{0}^{1} \mathbbm{1}\left(0 < \frac{1}{2\sigma_{Y}^{2}}\left\{\frac{\widetilde{Y}(\bld{s}_i) - \sum_{j:\bld{s}_j \in S_i} b_{ij}\widetilde{Y}(\bld{s}_j)}{\sqrt{f_i}}\right\}^{2} < -\log u(\bld{s}_i)\right) \nonumber \\
    &\qquad \mathbbm{1}(0 < u(\bld{s}_i) < 1) du(\bld{s}_i) \nonumber \\
    &= \exp\left(-\frac{1}{2\tau_{Y}^{2}}\widetilde{\bld{y}}'\widetilde{\bld{y}}\right) \prod_{i=1}^{n} \int_{0}^{1} \mathbbm{1}\left(0 < u(\bld{s}_i) < \exp \left(-\frac{1}{2\sigma_{Y}^{2}}\left\{\frac{\widetilde{Y}(\bld{s}_i) - \sum_{j:\bld{s}_j \in S_i} b_{ij}\widetilde{Y}(\bld{s}_j)}{\sqrt{f_i}}\right\}^{2}\right) < 1\right) du(\bld{s}_i) \nonumber \\
    &= \exp\left(-\frac{1}{2\tau_{Y}^{2}}\widetilde{\bld{y}}'\widetilde{\bld{y}}\right) \prod_{i=1}^{n} \exp \left(-\frac{1}{2\sigma_{Y}^{2}}\left\{\frac{\widetilde{Y}(\bld{s}_i) - \sum_{j:\bld{s}_j \in S_i} b_{ij}\widetilde{Y}(\bld{s}_j)}{\sqrt{f_i}}\right\}^{2}\right) \nonumber \\
    &= \exp\left(-\frac{1}{2\tau_{Y}^{2}}\widetilde{\bld{y}}'\widetilde{\bld{y}}\right) \exp \left(-\frac{1}{2\sigma_{Y}^{2}}\sum_{i=1}^n \left\{\frac{\widetilde{Y}(\bld{s}_i) - \sum_{j:\bld{s}_j \in S_i} b_{ij}\widetilde{Y}(\bld{s}_j)}{\sqrt{f_i}}\right\}^{2}\right) \nonumber \\
    &= \exp\left(-\frac{1}{2\tau_{Y}^{2}}\widetilde{\bld{y}}'\widetilde{\bld{y}}\right) \exp\left(- \frac{1}{2\sigma_{Y}^{2}}\left[\bld{F}^{-1/2}(\widetilde{\bld{y}}-\bld{B}\widetilde{\bld{y}})\right]'\left[\bld{F}^{-1/2}(\widetilde{\bld{y}}-\bld{B}\widetilde{\bld{y}})\right]\right) \nonumber \\
    &= \exp\left(-\frac{1}{2}\left[\frac{1}{\tau_{Y}^{2}}\widetilde{\bld{y}}'\widetilde{\bld{y}} + \frac{1}{\sigma_{Y}^{2}}(\widetilde{\bld{y}}-\bld{B}\widetilde{\bld{y}})'\bld{F}^{-1}(\widetilde{\bld{y}}-\bld{B}\widetilde{\bld{y}})\right]\right) \nonumber \\
    &= \exp\left(-\frac{1}{2}\left[\frac{1}{\tau_{Y}^{2}}\widetilde{\bld{y}}'\widetilde{\bld{y}} + \frac{1}{\sigma_{Y}^{2}}\widetilde{\bld{y}}'(\bld{I}-\bld{B})'\bld{F}^{-1}(\bld{I}-\bld{B})\widetilde{\bld{y}}\right]\right) \nonumber \\
    &= \exp\left(-\frac{1}{2}(\bld{y}-\bld{X\beta})'\left[\frac{1}{\tau_{Y}^{2}}\bld{I} + \frac{1}{\sigma_{Y}^{2}}(\bld{I}-\bld{B})'\bld{F}^{-1}(\bld{I}-\bld{B})\right](\bld{y}-\bld{X\beta})\right), \nonumber \\
    &\implies \bld{y}\vert \bld{\beta},\sigma_{Y}^{2},\tau_{Y}^{2} \sim \mathcal{N}\left(\bld{X\beta},\left[\frac{1}{\tau_{Y}^{2}}\bld{I} + \frac{1}{\sigma_{Y}^{2}}(\bld{I}-\bld{B})'\bld{F}^{-1}(\bld{I}-\bld{B})\right]^{-1}\right). \label{AA5}
\end{align}
}%
This concludes the the proof.

Notice that, again the precision-nugget is added in the precision matrix of $\bld{y}\vert \bld{\beta},\sigma_{Y}^{2},\tau_{Y}^{2}$ in \eqref{2eq:condY3}. Also, $S_i$, $\bld{B}$ and $\bld{F}$ have same definitions as it is presented for nearest-neighbor Gaussian process (NNGP) model in \citet{datta2016nngp}. This leads to the following two important corollaries from Proposition \ref{2prop:6}.

\begin{corollary} \label{2cor:6}
    When $\tau_{Y}^{2} = \infty$, the model in \eqref{2eq:condY3} becomes an NNGP model with no nugget/error variance.
\end{corollary}
\textit{Proof:} If we set $\tau_{Y}^{2} = \infty$ into \eqref{2eq:condY3}, we get $\bld{y}\vert \bld{\beta},\sigma_{Y}^{2} \sim \mathcal{N}\left(\bld{X\beta},\sigma_{Y}^{2}[(\bld{I}-\bld{B})'\bld{F}^{-1}(\bld{I}-\bld{B})]^{-1}\right)$. This is an NNGP model with mean $\bld{X\beta}$ and covariance matrix $\sigma_{Y}^{2}[(\bld{I}-\bld{B})'\bld{F}^{-1}(\bld{I}-\bld{B})]^{-1}$ when there is no nugget (refer to Section \ref{2subsec:reviewNNGP}; also see \citealp{datta2016nngp,datta2022nearest} for standard references). Note that, we consider $\bld{\mathcal{C}}$ to be a correlation function. This concludes the proof.

\begin{corollary} \label{2cor:7}
    The precision matrix $\frac{1}{\tau_{Y}^{2}}\bld{I} + \frac{1}{\sigma_{Y}^{2}}(\bld{I}-\bld{B})'\bld{F}^{-1}(\bld{I}-\bld{B})$ of the distribution in \eqref{2eq:condY3} is strictly positive definite.
\end{corollary}
\textit{Proof:} For any non-zero vector $\bld{v}$ we get $\bld{v}'\left[\frac{1}{\tau_{Y}^{2}}\bld{I} + \frac{1}{\sigma_{Y}^{2}}(\bld{I}-\bld{B})'\bld{F}^{-1}(\bld{I}-\bld{B})\right]\bld{v} = \sum_{i} v_{i}^{2}/\tau_{Y}^{2} + \sum_{i}q_{i}^{2}/\sigma_{Y}^{2}$, where the vector $\bld{q}=\bld{F}^{-1/2}(\bld{I}-\bld{B})\bld{v}$. Since $v_{i}^{2}>0$ and $q_{i}^{2}\geq 0$, $\forall i=1,\ldots,n$, we find $\bld{v}'\left[\frac{1}{\tau_{Y}^{2}}\bld{I} + \frac{1}{\sigma_{Y}^{2}}(\bld{I}-\bld{B})'\bld{F}^{-1}(\bld{I}-\bld{B})\right]\bld{v} > 0$. Therefore, $\frac{1}{\tau_{Y}^{2}}\bld{I} + \frac{1}{\sigma_{Y}^{2}}(\bld{I}-\bld{B})'\bld{F}^{-1}(\bld{I}-\bld{B})$ is strictly positive definite, and this concludes the proof.

Corollary \ref{2cor:6} shows that $\tau_{Y}^{2}$ makes this version of the TAR model an NNGP model with no precision-nugget. Hence, we can refer the model in \eqref{2eq:trunc3} as the NNGP version of the TAR model in the process setting. However, NNGP is an approximation method whereas the model in \eqref{2eq:trunc3} is a well defined Markov random field. Thus, if one believes that the process is actually distributed according to the covariance function that the Vecchia approximation is approximating then this TAR\textsubscript{S} model with $\tau_{Y}^{2}=\infty$ leading to the NNGP is preferable.

\section{Direct Sampler on CAR and SAR Model} \label{2app:c}
Here, we find the posterior distribution from the CAR and SAR model when implementing a direct sampler. We consider the hierarchical model
\begin{align}
    \bld{y}_{O}\vert\bld{\beta},\sigma_{Y}^{2},\rho &\sim \mathcal{N}\left(\bld{X}_{O}\bld{\beta},\sigma_{Y}^{2}\left(\bld{O}\bld{\Sigma}(\rho)^{-1}\bld{O}'\right)^{-1}\right), \nonumber \\
    \pi(\bld{\beta}) &= 1, \nonumber \\
    \sigma_{Y}^{2} &\sim \mathcal{IG}(a,b), \nonumber \\
    \rho &\sim \mathcal{DU}(\rho_{1},\ldots,\rho_{k}), \label{BB1}
\end{align}
where $\sigma_{Y}^{2}$ follows an inverse gamma prior with shape and scale parameter as $a$ and $b$ respectively, and $\rho$ follows a discrete uniform prior with support $\{\rho_{1},\ldots,\rho_{k}\}$. $\bld{\Sigma}(\rho) = \left(\bld{D}_{w}-\rho\bld{W}\right)^{-1}$ for a CAR model and  $\bld{\Sigma}(\rho) = \left[(\bld{I}-\rho\bld{A})'(\bld{I}-\rho\bld{A})\right]^{-1}$ for a SAR model. Our goal is to find the posterior distribution $f(\bld{\beta},\sigma_{Y}^{2},\rho\vert\bld{y}_{O})=f(\bld{\beta}\vert\sigma_{Y}^{2},\rho,\bld{y}_{O})f(\sigma_{Y}^{2}\vert\rho,\bld{y}_{O})f(\rho\vert\bld{y}_{O})$ by making use of a direct sampler. The derivation can be done by following exactly the same steps as given in Appendix \ref{2app:a3}, where substitute $\delta$ with $\rho$, $\bld{\Sigma}(\delta)$ with $\bld{\Sigma}(\rho)$, and $\{\delta_{1},\ldots,\delta_{k}\}$ with $\{\rho_{1},\ldots,\rho_{k}\}$. Also, the posterior predictive distribution of CAR and SAR follows \eqref{2eq:postpred} with the aforementioned substitutes.

\newpage
\bibliographystyle{apalike}
\bibliography{bibliography.bib}

\end{document}